\documentclass[11pt,a4paper]{article}

\usepackage{jheppub}
\usepackage[T1]{fontenc}
\usepackage{fix-cm}
\usepackage{lmodern}
\usepackage{mathtools}
\usepackage{mathrsfs}
\usepackage{empheq}
\usepackage{physics}
\usepackage{bbm}
\usepackage{bm}
\usepackage{hhline}
\usepackage{tensor}
\usepackage{tikz}
\usepackage{xcolor}
\usepackage{pifont}
\usepackage{latexsym}
\usepackage{slashed}
\usepackage{multirow}
\usepackage{float}
\usepackage{afterpage}

\makeatletter
\newcommand*{\rom}[1]{\expandafter\@slowromancap\romannumeral #1@}
\makeatother

\newcommand\cN{\mathcal{N}}

\newcommand\cR{\mathcal{R}}

\newcommand{\cA}{\mathcal{A}}
\newcommand{\cB}{\mathcal{B}}
\newcommand{\cF}{\mathcal{F}}
\newcommand{\cG}{\mathcal{G}}

\newcommand{\red}{\color{red}}
\newcommand{\SO}{{\rm SO}}
\newcommand{\Red}{\color{red}}
\newcommand{\DS}{Dynamical System }

\title{New $G_2$-invariant supersymmetric AdS$_4$ vacua in 11d supergravity and holography}
\author[a,b]{Xuao Zhang}

\affiliation[a]{Kavli Institute for Theoretical Sciences (KITS), \\
	University of Chinese Academy of Sciences, Beijing 100190, China}

\affiliation[b]{Instituut voor Theoretische Fysica, KU Leuven, \\
	Celestijnenlaan 200D, B-3001 Leuven, Belgium}

\emailAdd{zhangxuao@ucas.ac.cn}

\abstract{We study the most general $G_2$-invariant ${\rm AdS}_4$ vacua in 11 dimensional supergravity preserving 4 real supercharges, with the goal of understanding the IR fixed points of the RG flow induced by the cubic deformation of the ABJM theory. We identify a new $G_2$-invariant background, which completes the web of RG flows connecting its holographic dual with the ABJM theory and the (cubic analogue of) mass-deformed ABJM theory with SU(3)$\times$U(1) isometry. Our holographic study gives non-trivial predictions for the strongly-coupled field theory dual. }

\begin{document}
	
\maketitle

\section{Introduction}
\label{sec:intro}
	
	Non-trivial information of quantum gravity could be extracted from the brane worldvolume quantum field theories in superstring or M theory. A special case in M theory, for a stack of $N$ M2 branes probing the conical singularity of a Calabi-Yau four-fold over the Sasaki-Einstein manifold $S^7/\mathbb{Z}_k$, the worldvolume theory could be described by the ABJM theory \cite{Aharony:2008ug}. The theory is a 3d $\cN = 6$ Chern-Simons matter theory with level $k$. In our work, we will be interested in the special case $k = 1$, for which the supersymmetry is enhanced to $\cN = 8$ \cite{Gustavsson:2009pm}, the maximal supersymmetry of a 3d superconformal field theory \cite{Nahm:1977tg}. According to the standard prescription of the AdS/CFT correspondence \cite{Maldacena:1997re, Gubser:1998bc, Witten:1998qj, Aharony:1999ti}, the quantum gravity theory dual to the ABJM theory could be approximated by an 11 dimensional supergravity background ${\rm AlAdS}_4 \times S^7$ whose external space is asymptotically locally AdS$_4$ and internal space is the seven-sphere.
	
\begin{figure}[h]
		\centering
		\caption{Illustration for motivating the conjectured $G_2$ solution in 11d supergravity. The left-hand half in red is the web of RG flow induced by a quadratic superpotential deformation of the ABJM theory. \cite{Bobev:2009ms} The right-hand half in blue is the counterpart with a cubic superpotential deformation, where the $G_2$ solution in the dashed box and corresponding RG flows are not yet discussed in the literature.}
		\includegraphics[width=8cm]{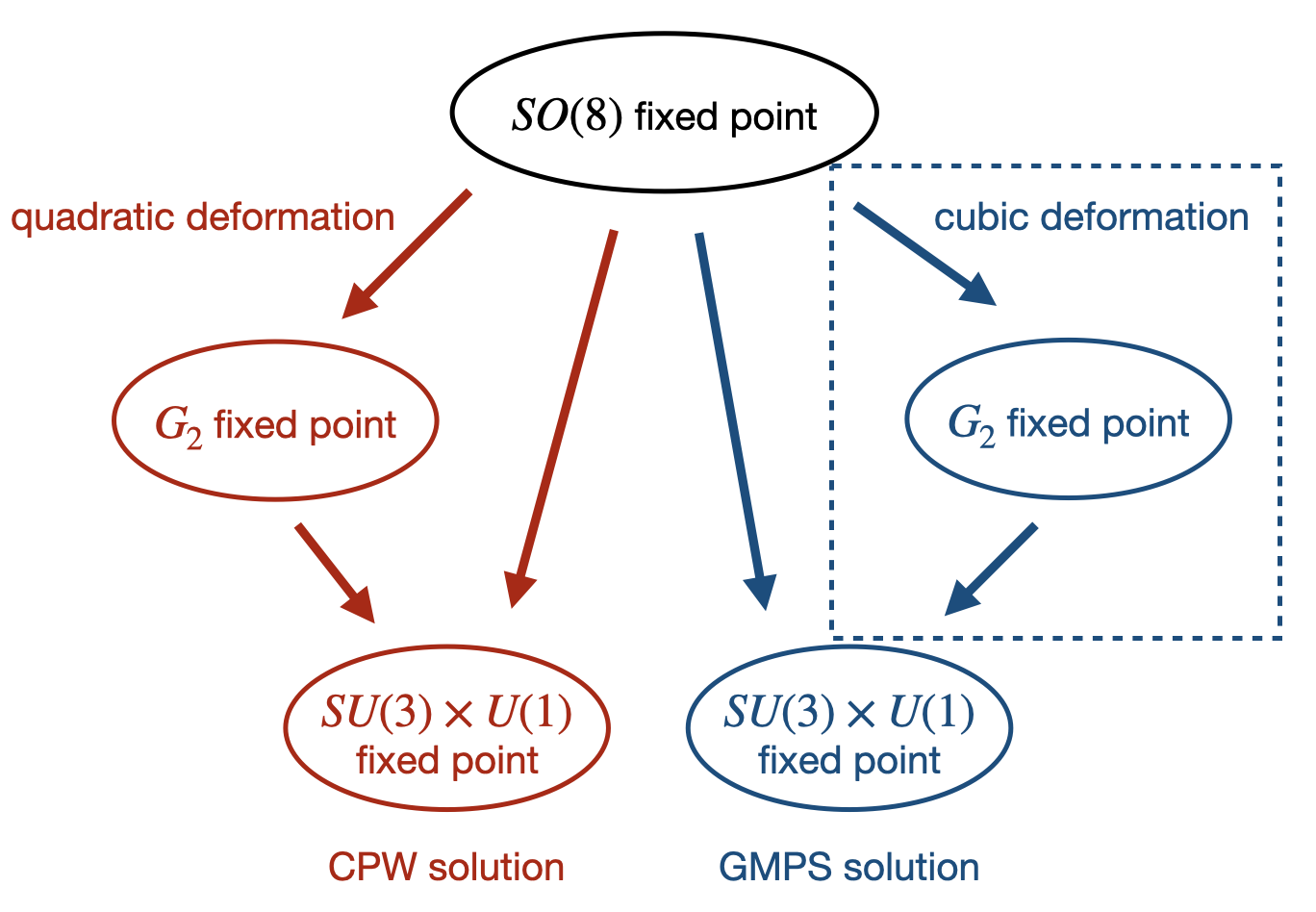}
		\label{fig:G2FixPt}
	\end{figure}

	Although ABJM theory is interesting on its own right, a rich structure could be probed by studying its deformations. As an example, for the maximally supersymmetric ABJM theory, one can turn on the scalar fields in the vector multiplets coupling to its global symmetry and breaks the theory down to $\cN = 2$. This deformation is called real mass deformation and can be studied by supersymmetric localisation. \cite{Freedman:2013oja} Here, we consider another type of deformation, where relevant deformations are added to the superpotential and induce renormalisation group flows. When the deformation is quadratic in the chiral superfields, it amounts to adding mass terms. For this case, the web of RG flows and IR fixed points have been discussed in \cite{Bobev:2009ms}, where there are two conformal fixed points: one is an $\cN = 1$ fixed point with $G_2$ global symmetry, and the other is an $\cN = 2$ fixed point with SU(3) flavor symmetry and U(1) $R$-symmetry, as shown in the red part of Fig. \ref{fig:G2FixPt}. The two fixed points have been identified with extrema of the scalar potential in 4d maximal SO(8) gauged supergravity \cite{Warner:1983vz}, and their uplifts to 11 dimensional supergravity are given by the ($G_2$-invariant) de Wit-Nocolai-Warner solution (dWNW) \cite{deWit:1984nz} and the (SU(3)$\times$U(1)-invariant) Corrado-Pilch-Warner solution (CPW) \cite{Corrado:2001nv}, respectively.
	
	Note that ABJM theory also admits relevant superpotential deformations that are cubic in the chiral superfields \cite{Jafferis:2011zi, Cordova:2016xhm}, these relevant deformations preserve $\cN = 1$ supersymmetry and will probably lead to IR fixed points. One fixed point is the 3d $\cN = 2$ theory with SU(3)$\times$U(1)$_R$ global symmetry whose holographic dual in 11 dimensional supergravity is found later in \cite{Gabella:2012rc} dubbed the Gabella-Martelli-Passias-Sparks (GMPS) solution (or HPW solution \cite{Halmagyi:2012ic}). Because of its enhanced 3d $\cN = 2$ supersymmetry, it could be studied from the field theory side by supersymmetric localisation \cite{Jafferis:2011zi}. The analogue between the two sides of Fig. \ref{fig:G2FixPt} motivates us to ask the following question: does the cubic deformation also lead to a 3d $\cN = 1$ $G_2$-invariant fixed point, which is connected to the other two fixed points by RG flows? A strong support for this proposal is given by the irreducible representation of the deformations under the global symmetry group. We find that the cubic single-trace deformation has a singlet sector under the $G_2$ group, as shown in Appendix \ref{App:G2singlet}, and thus may lead to an isometry-preserving RG flow and an IR fixed point. Because of the low amount of supersymmetry, we cannot use the powerful tool of supersymmetric localisation \cite{Pestun:2007rz, Pestun:2016zxk}. By virtue of the AdS/CFT correspondence, we can gain useful information of the conjectured fixed point by constructing and studying its supergravity dual.
	
	The 11 dimensional supergravity has a spontaneous compactification on $S^7$ which results in a consistent truncation \cite{deWit:1984nz, deWit:1986oxb, Nicolai:2011cy} (a classical review of the context is \cite{Duff:1986hr}) down to the 4 dimensional maximal supergravity with SO(8) gauging \cite{deWit:1982bul}. After making the truncation, the renormalisation group fixed points can be identified with extrema of the scalar potential in gauged supergravity. \cite{Warner:1983vz} Because the quadratic deformation itself belongs to the maximal SO(8) gauged supergravity, the holographic RG flows can be studied after the truncation, as nicely done in \cite{Bobev:2009ms}. However, the cubic deformation lies beyond the truncation \cite{DHoker:2000pvz}, meaning that we have to work in the original 11 dimensional supergravity. In this context, the $G_2$-invariant solutions have been discussed in the literature  \cite{Gunaydin:1983mi, deWit:1984nz} for some special cases, and very recently in \cite{Duboeuf:2024tbd} using exceptional field theory techniques. In our work, we study the most general $G_2$-invariant background in 11 dimensional supergravity preserving four supercharges and identify the new $G_2$ saddle as the bulk dual of the conjectured $G_2$-invariant fixed point. This complements the recent work \cite{Duboeuf:2024tbd}.
	
	Supersymmetric conditions have played an important role in identifying supersymmetric solutions in various supergravity theories, such as \cite{Corrado:2001nv, Pope:2003jp, Gowdigere:2003jf, Pilch:2003jg, Nemeschansky:2004yh} (see also \cite{Halmagyi:2012ic,Pilch:2015vha, Pilch:2015dwa} for more recent work). They are usually enbodied in the form of spinor projectors, which are idempotent operators that project out half of the components of the Killing spinors. By studying the Killing spinors on the known dWNW $G_2$-invariant background, we manage to extract the BPS conditions, which help us identify some new $G_2$-invariant backgrounds in 11 dimensional supergravity. 

	The structure of the paper is as follows: in section \ref{sec:eombps}, we study the Killing spinors in the known dWNW $G_2$-invariant solution and get the BPS equations. This section is independent on its own and can be skipped for people interested in the new saddle. Then in section \ref{sec:G2Saddles}, we treat the BPS equations as dynamical systems in a three-dimensional parameter space and show how the old and new $G_2$ saddles obtained from this picture. A complete search for solutions of the BPS equations is relegated to Appendix \ref{App:DynamicalSystem}. With the new solution being presented, in section \ref{sec:freeEnergy} we study the holographic free energy and provide evidence for the web of RG flow. Some questions for the future are discussed in section \ref{sec:ccl}.

	{\it Note added:} Recently, the interesting paper \cite{Duboeuf:2024tbd} also shows new $G_2$-invariant saddles. Compared to their method, we take advantage of supersymmetry and simplify the problem significantly. The solution denoted by G$_2'$ there is identical to the one we present in section \ref{subsec:IntegralCurvesBranch2}. With the BPS conditions, we can further show that the other $G_2$ saddle dubbed $G_2''$ is non-supersymmetric. We put the details of the comparison in Appendix \ref{App:CompareDuboeuf}.

\section{The equations of motion and BPS equations}
\label{sec:eombps}

	\subsection{The 11d supergravity and $G_2$-invariant ansatz}
	
	The 11d supergravity \cite{Cremmer:1978km} is the unique supergravity theory in 11 dimensional spacetime \cite{Nahm:1977tg}. Its matter contents include the graviton $g_{MN}$ and a 3-form gauge potential $A_{MNP}$ with field strength $F_{MNPQ}$. \footnote{Our convention for the space-time indices:
		\begin{itemize}
			\item $M, N, P, \cdots$ are curved incides on the 11d space-time that range from 1 to 11, and $A, B, C, \cdots$ are the corresponding flat indices.
			
			\item $\theta$ is the fifth coordinate in 11d, and $m,n,p,\cdots$ are curved indices on $S^6$ ranging from 6 to 11.
	\end{itemize} \label{footnote:indices} }
	
	\begin{equation} \label{eq:eom11d}
		\begin{aligned}
			& R_{MN} -\frac{1}{12}\left[ F_{MPQR}F_N^{\ \; PQR} - \frac{1}{12} g_{MN}F^2 \right] = 0, &\\
			& {\rm d}\star_{11} F + \frac{1}{2} F\wedge F = 0,\qquad F^2 \equiv F_{MNPQ}F^{MNPQ}. &\\
		\end{aligned}
	\end{equation}
	To clarify our convention of the differential geometry, we write down the flux equation in component form:
	\begin{equation}
		\nabla_M F^{MNPQ} + \frac{1}{2^7 3^2} \epsilon^{NPQM_1M_2M_3M_4M_5M_6M_7M_8}F_{M_1M_2M_3M_4}F_{M_5M_6M_7M_8} = 0.
	\end{equation}
	In 11d supergravity, the most general metric ansatz that preserves (at least) $G_2$ symmetry is as follows: \cite{Ahn:2001kw}
	\begin{equation} \label{eq:ansatzMetric}
		ds^2 = e^{2f_0(\theta)}ds^2_{\rm AdS_4} + e^{2f_1(\theta)}d\theta^2 + e^{2f_2(\theta)} d\Omega_6^2,
	\end{equation}
	where we denote $ds_{\rm AdS_4}$ and $d\Omega_6$ as the line element on AdS$_4$ and $S^6$ with unit radius. The range of $\theta$ is a finite interval $I_\theta$ that may change up to reparametrization, which allows us to choose $f_1(\theta)$ freely. The metric ansatz itself preserves SO(7) isometry, but the gauge field along the internal manifold further breaks it down to $G_2$:
	\begin{equation} \label{eq:ansatzA}
		A_{\theta mn} = g_1(\theta) J_{mn},\quad A_{mnp} = g_2(\theta) T_{mnp} + g_3(\theta) S_{mnp},
	\end{equation}
	where the space-time indices are explained in footnote \ref{footnote:indices}. The tensors $J_{mn}, T_{mnp}, S_{mnp}$ are the almost complex form, the torsion and dual torsion on $S^6$, they form the set of all possible $G_2$ invariant tensors on $S^6$  \cite{Gunaydin:1983mi}, whose constructions are shown explicitly in Appendix \ref{App:G2tensors}. As is explained in footnote 7 of \cite{Ahn:2001kw}, all the other components of $A_{MNP}$ vanish.
	
	The 4-form field strength is given by:
	\begin{equation}
	\begin{aligned}
				F^{(4)} &= g_0\, {\rm vol}_{\rm AdS_4} + {\rm d}A^{(3)},\\
	\end{aligned}
	\end{equation}
	 where $g_0$ is fixed to be a constant by the Bianchi identity. Plugging in the ansatz for the gauge potential \eqref{eq:ansatzA}, we get the components of the field strength:
	 \begin{equation} \label{eq:ansatzF}
		F_{\mu\nu\rho\sigma} = g_0 \epsilon_{\mu\nu\rho\sigma},\quad F_{\theta mnp} = (g_2'-3g_1)T_{mnp} + g_3'S_{mnp},\quad F_{mnpq} = 2g_3 \epsilon_{mnpqrs}J^{rs},
	 \end{equation}
	where the prime denotes derivative of $\theta$. Notice that the field strength only depends on the combination $g_2' - 3g_1$, this indicates the gauge redundancy in the ansatz of the gauge potential. 
	
	Plugging in the $G_2$-invariant ansatz \eqref{eq:ansatzMetric}\eqref{eq:ansatzF} into the equations of motion, we extract three independent differential equations from the Einstein equation:
	\begin{equation} \label{eq:eomComponents}
		\begin{aligned}
			& -f_0''+f_0' \left(-4 f_0'+f_1'-6 f_2'\right)+\frac{1}{3} g_0^2 \left(e^{6 f_2}+2 g_3^2\right) e^{-8 f_0+2 f_1-6 f_2}\\
			& \hspace{6cm} -3 e^{2 f_1-2 f_0}+8 g_3^2 e^{2 f_1-8 f_2}+\frac{2}{3}
			e^{-6 f_2} g_3'^2 = 0, \\
			& -f_2'' + f_2' \left(-4 f_0'+f_1'-6 f_2'\right)-\frac{1}{6} g_0^2 \left(e^{6 f_2}+2 g_3^2\right) e^{-8
				f_0+2 f_1-6 f_2}\\
				&  \hspace{6cm} -8 g_3^2 e^{2 f_1-8 f_2}+5 e^{2 f_1-2 f_2} -\frac{1}{3} e^{-6 f_2} g_3'^2 = 0, \\
			& 8 f_0' f_2'+2 f_0'^2-\frac{1}{12} g_0^2 \left(e^{6 f_2}+4 g_3^2\right) e^{-8 f_0+2 f_1-6
				f_2}\\
				&  \hspace{3cm} +e^{2 f_1} \left(2 e^{-2 f_0}-5 e^{-2 f_2}\right)+4 g_3^2 e^{2 f_1-8 f_2}+5 f_2'^2-\frac{1}{3} e^{-6 f_2} g_3'^2 = 0, \\
		\end{aligned}
	\end{equation}
	and another two differential and algebraic equations from the Maxwell equation:
	\begin{equation}
	\begin{aligned}
		&g_3'' + g_3' \left( 4f_0'-f_1'\right) + e^{2f_1}\left( -12 e^{-2f_2} + e^{-8f_0} g_0^2 \right) g_3 = 0, \\
		& \left( g_2' - 3g_1\right) = e^{f_1-4f_0} g_0 g_3. \\
		\end{aligned}
	\end{equation}
	\begin{table}[!h]
		\small 
		\centering
		\caption{\rm Comparison of conventions and notations of the bosonic fields. The presentation is such that expressions on the same line are equal to each other under different conventions. A typo in \cite{Gunaydin:1983mi} is highlighted in red. }
		\begin{tabular}{c|c|c|c|c|c|c}
			\hline 
			 & GMPS \cite{Gabella:2012rc} & \cite{Godazgar:2013nma} & dWNW \cite{deWit:1984nz, deWit:1986mz} & GW \cite{Gunaydin:1983mi} &  \cite{Ahn:2001kw} & CPW \cite{Corrado:2001nv} \\ \hline
			$e^{f_0(\theta)}$ & $\frac{1}{2}e^\Delta$ (2.3) & $H^{C_4}(\varphi)$ & $\sigma(\theta)$(4.1) & $1$  && $H^{1/3}(\rho) $ \\
			$e^{f_1(\theta)}$ & & $H^{C_7}(\varphi)$ & $\rho_1(\theta) $ & $1$ &&\\
			$e^{f_2(\theta)}$ & & $\rho H^{C_7}(\varphi)$ & $\rho_2(\theta)\sin\theta $ & $\rho(\alpha)$(3.3) &&\\
			\hline 
			$F$ & $G$ &  $ -\sqrt{2}iF $ &  $ -\sqrt{2}iF $ & $2F$ & $2F$ & $-2F$ \\
			$\frac{F_{1234}}{\epsilon_{1234}}=g_0$ & $2^{-4}m$ & $im$ & $if$(3.3) & $-2m$(3.2) &&\\
			$g_0$ & $2^{-4}m$ & $\sqrt{2}m$ & $\sqrt{2}f$ & $-\textcolor{red}{8}m$ &&\\
			$g_1(\theta)$ & & $\sqrt{2}h(\varphi)$ && $2h$ &&\\
			$g_2(\theta)$ & & $\sqrt{2}f_1(\varphi)$ && $2f_1$& &\\
			$g_3(\theta)$ & & $\sqrt{2}f_2(\varphi)$ & $\sqrt{2}f_1(\theta)$ & $2f_2$ &&\\ \hline 
		\end{tabular}
		\label{tbl:ComparisonConventions}
	\end{table}
	To make sure our results are consistent with the literature, we compare our convention with others in Table \ref{tbl:ComparisonConventions}. For example, we can identify our notation with that of \cite{deWit:1984nz}:
	\begin{equation}
		\begin{aligned}
		& e^{2f_0} = \sigma^2(\theta),\quad e^{2f_1} = \rho_1^2(\theta),\quad e^{2f_2} = \rho_2^2\sin^2\theta,\\
		& g_3 = \sqrt{2} f_1,\quad g_2'-3g_1 = -\sqrt{2} f_2,\quad g_3' = -\sqrt{2} f_3,\quad g_0 = \sqrt{2} f.
	\end{aligned}\label{eq:ConventiondWNW1985}
	\end{equation}
	For future convenience, let's review the known solution in the literature with SO(8) and $G_2$ \cite{deWit:1984nz} isometry using our notation:
	\begin{equation}	\label{eq:AdS4S7}
		{\rm SO(8):} \quad e^{f_0} = \frac{g_0^{1/3}}{3^{1/3}},\quad e^{f_1} =  \frac{2 g_0^{1/3}}{3^{1/3}},\quad e^{f_2} = \frac{2 g_0^{1/3}}{3^{1/3}} \sin\theta,\quad (g_2' - 3g_1) = g_3 = 0.
	\end{equation}
	\begin{equation} \label{eq:dWNWSolution}
	 \begin{aligned} 
		& G_2: \quad e^{f_0} = \frac{2^{\frac{1}{6}}}{3^{\frac{1}{6}} 5^{\frac{1}{6}}}  g_0^{\frac{1}{3}} \left(2+\cos2\theta\right)^{\frac{1}{3}},\quad e^{f_1} = \frac{2^{\frac{5}{3} }}{3^{\frac{1}{6}} 5^{\frac{2}{3}} } g_0^{\frac{1}{3}} \left(2+\cos2\theta\right)^{\frac{1}{3}}, \\
		& \quad e^{f_2} = \frac{2^{\frac{5}{3}}3^{\frac{1}{3}}}{5^{\frac{2}{3}}}  g_0^{\frac{1}{3}} \sin\theta  \left(2+\cos2\theta\right)^{-\frac{1}{6}}, \\
		&\quad g_2'-3g_1 = \frac{2^6 3^{\frac{3}{2}}}{5^{\frac{5}{2}}} g_0 \sin^4\theta \left(2+\cos2\theta\right)^{-2},\quad g_3 = \frac{2^5 3^1}{5^{\frac{5}{2}}} g_0 \sin^4\theta \left(2+\cos2\theta\right)^{-1}. \\
	\end{aligned}
	\end{equation}
	Notice that the family of solutions is invariant under the trombone symmetry \cite{Cremmer:1997xj} corresponding to the rescaling with a positive $\lambda$:
	\begin{equation} \label{eq:trombone}
	e^{f_i}\rightarrow \lambda e^{f_i}; \quad g_j\rightarrow \lambda^3 g_j.\quad i = 1,2,3;\ j = 0,1,2,3.
	\end{equation}
	The symmetry changes the size of the external AdS$_4$ and internal compact space, but it will not affect the holographic free energy we study in section \ref{sec:freeEnergy}. Thus, in practice we can always choose a suitable $\lambda$ so that $g_0 = 1$ without affecting the physics.
	
	\subsection{The supersymmetric variations}
		
	Second order defferential equations like the equations of motion are hard to solve in general. By imposing the supersymmetry conditions, one gets the BPS equations which only involve first order derivatives and are much easier to deal with both analytically and numerically. Our goal here is to study the supersymmetric conditions in a general $G_2$ background.
	
	On a supersymmetric background, we expect the supersymmetric variations of all the fields, i.e., the graviton, the Rarita-Schwinger field, and the 3-form gauge potential to vanish. The variations of the bosons vanish trivially because the fermion is turned off on the background, thus the only non-trivial constraint comes from the variation of the Rarita-Schwinger field:
	\begin{equation} \label{eq:susyVariation}
		\begin{aligned}
			& 	\delta \Psi_M = \nabla_M \epsilon + \frac{1}{288}\left(\Gamma_M^{\ \  N_1N_2N_3N_4}-8\delta_M^{N_1}\Gamma^{N_2N_3N_4}\right) F_{N_1N_2N_3N_4}\epsilon = 0, \\
			& \nabla_M \epsilon \equiv \partial_M \epsilon + \frac{1}{4}\omega_M^{\ \ AB}\Gamma_{AB}\epsilon,\quad \omega_M^{\ \ AB} \equiv E_N^A \bar{E}^{BL} \Gamma^N_{ML} - \bar{E}^{LB} \partial_M E^A_L, \\
		\end{aligned}
	\end{equation}
	where the space-time indices are explained in footnote \ref{footnote:indices}. In the last line, $\Gamma^N_{ML}$ denotes the Levi-Civita connection and shouldn't be confused with the Gamma-matrices. If there exists a non-zero variation generated by a spinor $\epsilon$ such that $\delta \Psi_M= 0$, the background preserves supersymmetry, and $\epsilon$ is the associated Killing spinor. Essentially, the supersymmetric constraints (such as spinor projectors) are the conditions on the metric and gauge field to guarantee the existence of Killing spinors.
	
	According to the standard behavior of the supersymmetric background, we expect half of the Killing spinors to be translational invariant on the plane $(t, x, y)$, and another half to have non-trivial spatial dependence. (see e.g. \cite{Pestun:2007rz}) The two sets of Killing spinors correspond to the $Q$-charges and $S$-charges in the dual superconformal field theory, respectively. For simplicity, we only consider the first set of Killing spinors in the following discussion. Plugging in the $G_2$ invariant ansatz \eqref{eq:ansatzMetric}\eqref{eq:ansatzF}, the supersymmetric variation of the gravitino gives: \footnote{Here the \textcolor{red}{red} indices are vielbein indices, and the black ones are the space-time indices. Our ordering of AdS$_4$ coordinates is such that $(x^1, x^2, x^3, x^4) = (r, t, x, y)$. The fifth coordinate is $\theta$ and the coordinates on $S^6$ range from 6 to 11.  }
	\begin{equation} \label{eq:KSE}
	\begin{aligned}
		& E_{\textcolor{red}{1}}^1  \delta\Psi_1 = \left( \partial_{\textcolor{red}{1}} + \frac{1}{2}e^{-f_1}f_0'\Gamma^{\textcolor{red}{15}} - \frac{1}{6}e^{-4f_0}g_0\Gamma^{\textcolor{red}{234}} +  \frac{1}{72}e^{-f_1}\Gamma^{ \textcolor{red}{15} mnp} F_{\theta mnp}  + \frac{1}{288}\Gamma^{\textcolor{red}{1}mnpq}F_{mnpq} \right) \epsilon, &\\
		& E_{\textcolor{red}{2}}^2   \delta\Psi_2 = \left(\frac{1}{2}e^{-f_0}\Gamma^{\textcolor{red}{12}} - \frac{1}{2}f_0'e^{-f_1}\Gamma^{\textcolor{red}{25}} + \frac{1}{6} g_0 e^{-4f_0}\Gamma^{\textcolor{red}{134}} +  \frac{1}{72}e^{-f_1}\Gamma^{ \textcolor{red}{25} mnp} F_{\theta mnp}  + \frac{1}{288}\Gamma^{\textcolor{red}{2}mnpq}F_{mnpq} \right)\epsilon, &\\
		& E_{\textcolor{red}{3}}^3  \delta \Psi_3 = \left( -\frac{1}{2}e^{-f_0}\Gamma^{\textcolor{red}{13}} + \frac{1}{2}f_0' e^{-f_1}\Gamma^{\textcolor{red}{35}} - \frac{1}{6} e^{-4f_0}g_0\Gamma^{\textcolor{red}{124}} +  \frac{1}{72}e^{-f_1}\Gamma^{ \textcolor{red}{35} mnp} F_{\theta mnp}  + \frac{1}{288}\Gamma^{\textcolor{red}{3}mnpq}F_{mnpq} \right) \epsilon, &\\
		& E_{\textcolor{red}{4}}^4  \delta\Psi_4 = \left( -\frac{1}{2}e^{-f_0} \Gamma^{\textcolor{red}{14}} + \frac{1}{2} e^{-f_1}f_0' \Gamma^{\textcolor{red}{45}} + \frac{1}{6}e^{-4f_0}g_0\Gamma^{\textcolor{red}{123}} + \frac{1}{72}e^{-f_1}\Gamma^{ \textcolor{red}{45} mnp} F_{\theta mnp}  + \frac{1}{288}\Gamma^{\textcolor{red}{4}mnpq}F_{mnpq} \right) \epsilon, &\\
		& E_{\textcolor{red}{5}}^5  \delta\Psi_5 = \left(\partial_{\textcolor{red}{5}} -\frac{1}{36}e^{-f_1}\Gamma^{mnp} F_{\theta mnp}  + \frac{1}{12}g_0\Gamma^{\textcolor{red}{12345}}e^{-4f_0} + \frac{1}{288} \Gamma_{\textcolor{red}{5}}^{\ \ mnpq} F_{mnpq}\right) \epsilon, &\\
		& E_{\textcolor{red}{6}}^6  \delta\Psi_6 = \left( \partial_{\textcolor{red}{6}} - \frac{1}{2}f_2'e^{-f_1}\Gamma^{\textcolor{red}{56}} + \frac{1}{12}e^{-4f_0}g_0\Gamma^{\textcolor{red}{12346}} + \frac{1}{12}e^{-f_1}\Gamma^{\textcolor{red}{5}mn} F_{\theta \textcolor{red}{6}mn} - \frac{1}{72}e^{-f_1}\Gamma^{\textcolor{red}{56}mnp} F_{\theta mnp}  \right. &\\
		&\left. \qquad + \frac{1}{288}\Gamma^{\textcolor{red}{6}mnpq}F_{mnpq} - \frac{1}{36}\Gamma^{mnp}F_{\textcolor{red}{6}mnp} \right)\epsilon, &\\
		&\cdots. &\\
	\end{aligned}
	\end{equation}
	The variations can be written in the following unified form:
	\begin{equation} \label{eq:KSECompactForm}
		E_{\textcolor{red}{M}}^M \delta \Psi_M = (\partial_{\textcolor{red}{M}} + P_{\textcolor{red}{M}}) \epsilon.
	\end{equation}
	We look at the variations of $t, x, y$ components first, which are algebraic equations of $\epsilon$ because of the translational invariance. We contract them with a Gamma matrix using the commutation relation \eqref{eq:GammaContraction} and find:
	\begin{equation}
		 \Gamma^{\textcolor{red}{12}} P_{\textcolor{red}{2}} \epsilon = \Gamma^{\textcolor{red}{13}} P_{\textcolor{red}{3}}  \epsilon = \Gamma^{\textcolor{red}{14}}P_{\textcolor{red}{4}}  \epsilon  = \left( \frac{1}{2} e^{-f_0} + P_{\textcolor{red}{1}} \right) \epsilon = 0.
	\end{equation}
	So the supersymmetric condition $\delta \Psi_M = 0$ with $M = 2,3,4$ dictates that
	\begin{equation} \label{eq:AlgebraicKSE}
		P_{\textcolor{red}{1}} \epsilon = - \frac{1}{2} e^{-f_0} \epsilon.
	\end{equation}
	Plugging it into the condition $ \delta \Psi_{\textcolor{red}{1}}  = 0$, we get:
	\begin{equation}
		\delta \Psi_{\textcolor{red}{1}} = (E_{\textcolor{red}{1}}^1 \partial_r + P_{\textcolor{red}{1}} ) \epsilon =  \left( e^{-f_0} \partial_r - \frac{1}{2} e^{-f_0} \right)\epsilon = 0 \quad \Rightarrow \quad \epsilon = e^{r/2} \epsilon_{\hat{r}},
	\end{equation}
	with $\epsilon_{\hat{r}}$ independent of $r$. By now we have completely determined the dependence of the Killing spinors on the external AdS$_4$, which holds true for all $G_2$ invariant supersymmetric backgrounds. So far our discussions have been general. The rest of the Killing spinor equations, however, involve differentials over the internal coordinates and are hard to solve analytically for general backgrounds. In what follows, we will study the known $G_2$ invariant dWNW solution \eqref{eq:dWNWSolution}, solve the algebraic Killing spinor equation (KSE) \eqref{eq:AlgebraicKSE} numerically and study the dependence of the Killing spinors on the internal space. The reason for doing this analysis is our assumption that these properties for Killing spinors in the dWNW background are shared by all $G_2$ invariant solutions. This assumption turns out to be valid and helps us obtain the simplified supersymmetric constraints.
	
	\subsection{The Killing spinors for the dWNW saddle}
	
	Now we move on to determine the dependence of the Killing spinors $\epsilon_{\hat{r}}$ on the internal space parametrised by $\psi_{i = 1,2,\cdots, 6}$ and $\theta$ by numerically solving the algebraic KSE \eqref{eq:AlgebraicKSE}, which can be written more explicitly as:\footnote{In practice, we use the {\tt Eigenvectors[]} function of Mathematica and take the eigenvectors whose eigenvalue is zero. }
	\begin{equation} \label{eq:AlgebraicKSEFull}
		\left( \frac{1}{3}g_0 e^{-3f_0} \Gamma^{\textcolor{red}{234} }  - f_0'e^{f_0 - f_1} \Gamma^{\textcolor{red}{15}} - \frac{1}{36} e^{f_0-f_1} \Gamma^{\textcolor{red}{15}mnp} F_{\theta mnp} - \frac{1}{144} e^{f_0} \Gamma^{\textcolor{red}{1}mnpq}F_{mnpq}\right) \epsilon_{\hat{r}} = \epsilon_{\hat{r}}.
	\end{equation}
	Following similar ansatz in \cite{Gabella:2012rc, Gauntlett:2005ww}, our spinor may have the following form:
	\begin{equation} \label{eq:KSCompactForm}
		\epsilon = \epsilon_{(4)} \otimes \theta \otimes  \epsilon_{(6)} +  \epsilon_{(4)}^c \otimes \theta \otimes  \epsilon_{(6)}^c,
	\end{equation}
	where $\epsilon_{(4)}$ and $ \epsilon_{(6)}$ are Killing spinors on AdS$_4$ and $S^6$ respectively. The factorization suggests us to use a specific representation of Gamma-matrices, where we follow the construction of \cite{Gauntlett:2004zh}. We admit the following representation of gamma matrices on $S^6$: \footnote{The indices are vielbein incides in the definiton of Gamma matrices below. }
	\begin{equation} \small 
		\begin{aligned}
			\gamma^1 &= \sigma_1 \otimes \mathbf{1} \otimes \mathbf{1},\quad \gamma^2 = \sigma_2 \otimes \mathbf{1} \otimes \mathbf{1}, \quad \gamma^3 = \sigma_3 \otimes \sigma_1 \otimes \mathbf{1}, \quad \gamma^4 = \sigma_3 \otimes \sigma_2 \otimes \mathbf{1}, \\
			\gamma^5 &= \sigma_3 \otimes \sigma_3 \otimes \mathbf{1},\quad \gamma^6 = \sigma_3 \otimes \sigma_3 \otimes \sigma_2,\quad \gamma^7 = -i \sigma_3 \otimes \sigma_3 \otimes \sigma_3,\\
		\end{aligned}
	\end{equation}
	where $\sigma_{i = 1,2,3}$ are Pauli matrices. They satisfy:
	\begin{equation} 
		\begin{aligned}
			&  (\gamma^m)^\dagger = \gamma^m; \qquad \{ \gamma^m, \gamma^n \} = 2 \delta^{mn},\quad m,n = 1,2,\cdots, 6;\\
			& (\gamma^7)^2 = - \mathbf{1},\quad \gamma^7 = \gamma^1\cdots \gamma^6.\\
		\end{aligned}
	\end{equation}
	We define the gamma matrices on ${\rm AdS}_4 \times I_\theta$ differently from usual with an extra factor of $i$:
	\begin{equation}
		\begin{aligned}
			\rho^1 & = i \sigma_1 \otimes \mathbf{1},\quad \rho^2 = - \sigma_2 \otimes \mathbf{1},\quad 
			\rho^3 = i \sigma_3 \otimes \sigma_1 ,\quad \rho^4 = i \sigma_3 \otimes \sigma_2, \quad
			\rho^5 = i \sigma_3 \otimes \sigma_3,\\
		\end{aligned}
	\end{equation}
	such that they satisfy\footnote{ This is different from \cite{Gauntlett:2004zh}, where they have $\rho^1 \rho^2\cdots \rho^5 = - \mathbf{1}$.   }
	\begin{equation}
		\begin{aligned}
			& \{\rho^a, \rho^b \} = \textcolor{red}{-} 2 \eta^{ab},\quad a,b = 1,2,\cdots , 5,\quad \eta^{ab} = {\rm diag}(1,-1,1,1,1), \quad 
		  \rho^1\cdots \rho^5 = + \mathbf{1}. \\
		\end{aligned}
	\end{equation}
	The 11d Gamma matrices are constructed in terms of them:
	\begin{equation}
		\begin{aligned}
			\Gamma^a &= \rho^a \otimes \gamma^7,\quad a = 1,2,\cdots, 5;\\
			\Gamma^{5+m} &= \mathbf{1}_{4\times 4} \otimes \gamma^m,\quad m = 1,2,\cdots, 6,\\
		\end{aligned}
	\end{equation}
	with the normal commutation relation and highest rank matrix given by:
	\begin{equation} \label{eq:GammaContraction}
		\{ 	\Gamma^{A} , \Gamma^B \} = 2 \eta^{AB},\quad \Gamma_* = \Gamma^1 \cdots \Gamma^{11} = - \mathbf{1}.
	\end{equation}
	One thing to notice before we continue is the homogeneity of the equation \eqref{eq:AlgebraicKSEFull}, which means any given solution can be multiplied by a complex factor $e^\xi e^{i\varphi}$ and produce another solution. To get a numerical solution, we have to fix both the normalisation factor $e^\xi$ and the phase factor $\varphi$ in the ambiguity. When numerically solving the KSE, we choose $\epsilon_{\rm num}^\dagger \epsilon_{\rm num} = 1$ for the overall normalisation to fix $\xi$ and require the first non-zero component to be real to fix the overall phase factor $\varphi$. However, the na\"ive way to fix $\xi$ and $\varphi$ turns out to be overconstraining, and we have to distinguish what we obtain numerically $\epsilon_{\rm num}$ and the correct candidates of Killing spinors $\epsilon_{\hat{r}}$:
	\begin{equation} \label{eq:ambiguousFactor}
		\epsilon_{\rm num} = e^\xi e^{i\varphi} \epsilon_{\hat{r}}.
	\end{equation}
	Let's demonstrate using coordinate $\psi_1$ how to fix these factors. Now we understand that it is $\epsilon_{\hat{r}}$ that solves the KSE \eqref{eq:KSECompactForm}, and the correct equation to be satisfied by the numerical spinors is actually:
	\begin{equation}
		(D_{\textcolor{red}{6}} + P_{\textcolor{red}{6}}) \epsilon_{\rm num} = 0,\quad D_{\textcolor{red}{6}} \equiv \partial_{\textcolor{red}{6}} -  i E_{\textcolor{red}{6}}^6 \varphi'(\psi_1)  - E_{\textcolor{red}{6}}^6 \xi'(\psi_1),
	\end{equation}
	where the difference exists in terms of $\varphi'$ and $\xi'$. The real and imaginary parts of the equation are, respectively: \footnote{We hide the suffix of $\epsilon_{\rm num}$ for simplicity. }
	\begin{equation}
		 \partial_{\textcolor{red}{6}} \epsilon^R + E_{\textcolor{red}{6}}^6 \varphi' \epsilon^I - E_{\textcolor{red}{6}}^6 \xi' \epsilon^R = - (P_{\textcolor{red}{6}} \epsilon)^R ,\quad  \partial_{\textcolor{red}{6}} \epsilon^I - E_{\textcolor{red}{6}}^6 \varphi' \epsilon^R - E_{\textcolor{red}{6}}^6 \xi'\epsilon^I = -(P_{\textcolor{red}{6}} \epsilon)^I,
	\end{equation}
	where the indices $R$ and $I$ denote the real and imaginary parts. By solving the two equations, we can fix the factors:
	\begin{equation} \label{eq:phipxip}
		E_{\textcolor{red}{6}}^6 \varphi' = \frac{-(P_{\textcolor{red}{6}} \epsilon)^R + (P_{\textcolor{red}{6}} \epsilon)^I + \partial_{\textcolor{red}{6}} \left( \epsilon^I - \epsilon^R \right) } { \epsilon^R + \epsilon^I } ,\quad E_{\textcolor{red}{6}}^6 \xi' = \frac{ \epsilon^I \left( \partial_{\textcolor{red}{6}} \epsilon^I +  (P_{\textcolor{red}{6}} \epsilon)^I  \right) + \epsilon^R \left( \partial_{\textcolor{red}{6}} \epsilon^R + (P_{\textcolor{red}{6}} \epsilon)^R \right) }{ \epsilon^R \left( \epsilon^R + \epsilon^I \right)  }.
	\end{equation}
	
	\begin{figure}[!h]
		\centering \caption{\rm The first and the second lines give $\varphi'(\psi_1)$ and $\xi'(\psi_1)$ evaluated through \eqref{eq:phipxip} for the numerical solution of the algebraic equation \eqref{eq:ambiguousFactor}. We have chosen components $1,3,4$, and 8 of $\epsilon_{\rm num}$ for $\psi_1 \in [0, 4\pi]$. Note that there are values of $\psi_1$ where certain components give a divergent $\varphi'$ or $\xi'$, this is because the right-hand side of \eqref{eq:phipxip} evaluates $0/0$ and becomes unstable numerically. In the second line, it looks like $\xi'$ obtained from different components are different, but the fact is that $\xi'$ is identically zero, and the non-zero fluctuations are all attributed to the artifect we mentioned above.  }
		\includegraphics[width=3.5cm]{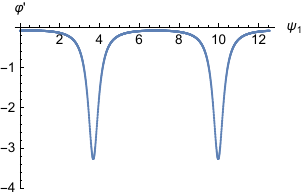}
		\includegraphics[width=3.5cm]{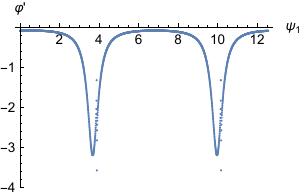}
		\includegraphics[width=3.5cm]{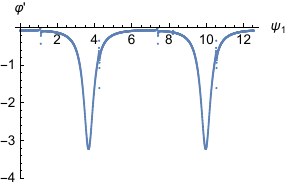}
		\includegraphics[width=3.5cm]{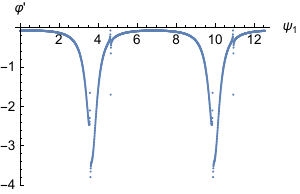} \\
		\includegraphics[width=3.5cm]{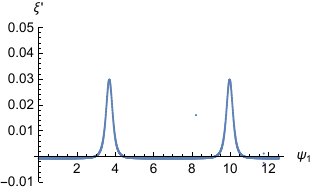}
		\includegraphics[width=3.5cm]{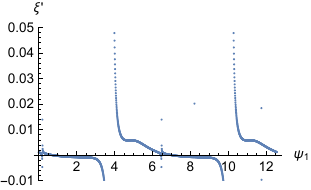}
		\includegraphics[width=3.5cm]{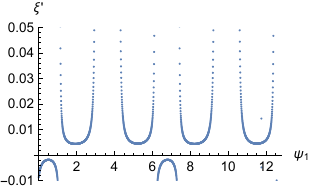}
		\includegraphics[width=3.5cm]{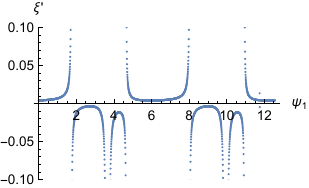}
		\label{fig:xiphi}
	\end{figure}
	\begin{figure}[!h]
		\centering \caption{\rm The blue dots show real factor $\xi'(\theta)$ evaluated numerically through \eqref{eq:phipxip} for $\theta \in [-\pi, \pi]$. We choose the result obtained from the first component of $\epsilon_{\rm num}$. The red curve is $-\frac{1}{2}f_0'(\theta)$ which is known analytically for the dWNW background \eqref{eq:dWNWSolution}. They match pretty well for most of the values. }
		\includegraphics[width=5cm]{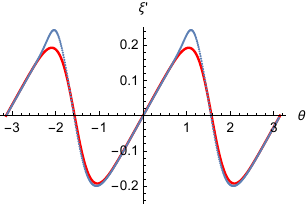}
		\label{fig:xiTheta}
	\end{figure}
	
	We need to check explicitly that all 32 components of $\epsilon_{\rm num}$ give the same $\varphi'$ and $\xi'$, since they are overall factors and have to be the same for all components. We have shown this in Fig. \ref{fig:xiphi}. We also observe that for $\psi_i$, the normalisation factor is automatically the ``correct'' one, i.e., $\xi' = 0$. This is also the case for all the coordinates on $S^6$, thus the normalisation is fixed by studying the $\theta$-dependence.
	
	For the variable $\theta$, we numerically verify that the normalisation factor $\xi(\theta)$ is related to the emblackening factor $f_0(\theta)$ of the dWNW metric, shown in Fig. \ref{fig:xiTheta}: \footnote{The functions $\xi(\theta)$ is a different function from $\xi(\psi_1)$ we use before, although we abuse the same notation for them. }
	\begin{equation}
		\xi'(\theta) = - \frac{1}{2} f'_0(\theta) = \frac{\sin(2\theta)}{3(2 + \cos(2\theta))}\quad \Rightarrow \quad \xi(\theta) = - \frac{1}{2} f_0(\theta) + {\rm const}.
	\end{equation}
	Plugging back to \eqref{eq:ambiguousFactor}, we get the following expression for the Killing spinor:
	\begin{equation}
		\epsilon = e^{r/2} \epsilon_{\hat{r}} = e^{r/2} e^{f_0/2} \epsilon_{ \hat{r}}^{\rm nor},
	\end{equation}
	such that $ (\epsilon_{ \hat{r}}^{\rm nor})^\dagger \epsilon_{ \hat{r}}^{\rm nor} \equiv 1$. In fact, we expect the Killing spinor to be expressed as a normalised constant spinor $\epsilon_{\rm const}^{\rm nor}$ rotated by non-Abelian factors such that: \cite{Gowdigere:2003jf}
	\begin{equation}
		\epsilon = e^{r/2} e^{f_0/2} \cR(\theta, \psi_i, \Gamma) \epsilon_{\rm const}^{\rm nor},
	\end{equation}
	with a complicated rotation matrix $\cR$ involving the internal coordinates and the Gamma matrices.
	
\begin{figure}
		\centering \caption{\rm We showcase the real and imaginary parts of the first eight components of $\epsilon_{\hat{r}}$ for $\psi_1 \in [0, 4\pi]$ for randomly chosen other coordinates. It's easy to see that they are periodic functions with period $4\pi$. }
		\includegraphics[width=16cm]{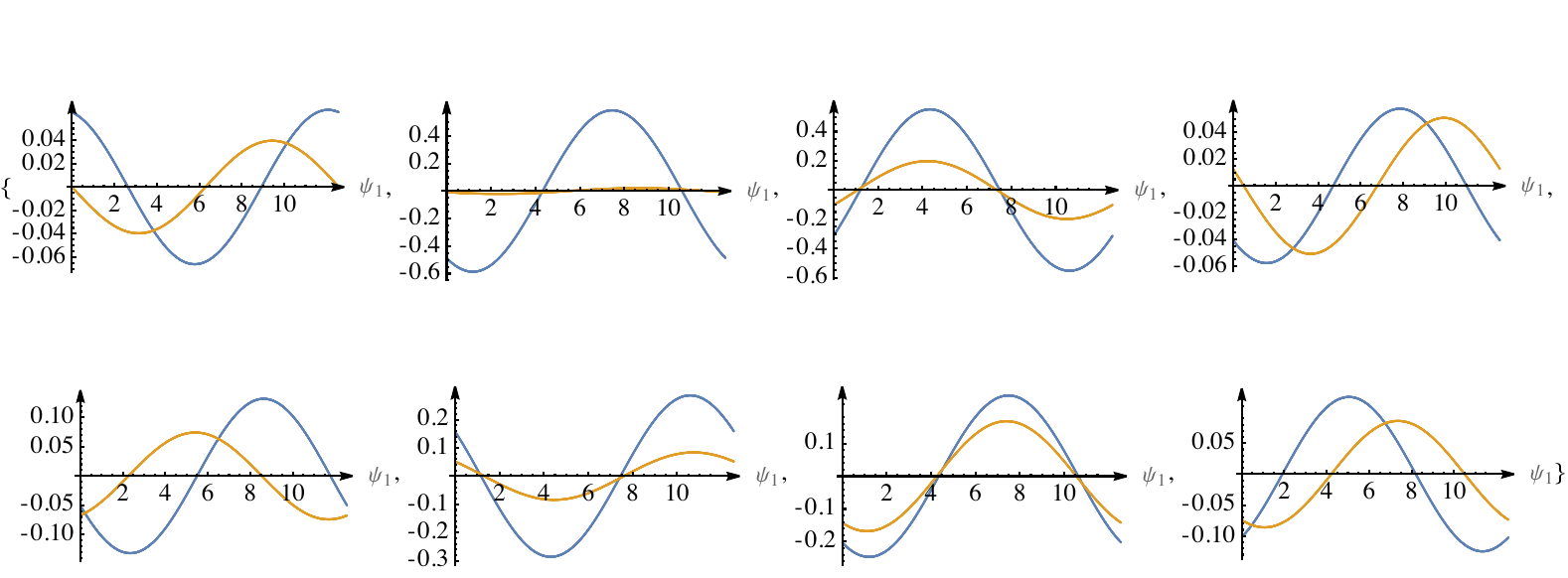}
		\label{fig:epsPsi1}
	\end{figure}

	Knowing how to determine the correct normalisation and phase factor, now we start solving the algebraic KSE numerically, this gives us two independent solutions, which we call $\epsilon_{\hat{r}}$ and $\eta_{\hat{r}}$.\footnote{From now on, we will only talk about the correct solutions of KSE instead of the ostensible one denoted by $\epsilon_{\rm num}$ above.} With our specific representation of Gamma-matrices, each of them have 16 out of 32 non-zero components and are related by:
		\begin{equation}
			\eta_{\hat{r}} = e^{i\delta } \Gamma^{\textcolor{red}{23}} \epsilon_{\hat{r}},
		\end{equation}
	where $\delta$ is a constant real number coming from the phase ambiguity we discussed above, which can be fixed to zero with a judicious gauge choice For what follows we can simply focus on $\epsilon$. Numerically, we find $\epsilon_{\hat{r}}$ also satisfy the charge-conjugation relation:
	\begin{equation}
		\epsilon_{\hat{r}} = e^{i\delta'} \Gamma^{\textcolor{red}{156810}} \epsilon^*_{\hat{r}},
	\end{equation}
	where $\delta'$ is another phase ambiguity that can be fixed to zero. By numerically fixing all other coordinates and varying the value of one coordinate $\psi_i$, we find that each component of the spinor $\epsilon_{\hat{r}}$ has the following dependence on $\psi_i$ (see Fig. \ref{fig:epsPsi1} as an example):
		\begin{equation}
		\begin{aligned}
			& 	\epsilon_{\hat{r}, \alpha} = c_{\hat{r}, \alpha} \cos \frac{\psi_i}{2} + s_{\hat{r}, \alpha} \sin \frac{\psi_i}{2}, \quad \alpha = 1,2,\cdots, 32.\\
		\end{aligned}
	\end{equation}
	This is reminiscent of the behavior of Killing spinors on $S^6$: \cite{Lu:1998nu}
	\begin{equation}  \label{eq:KSonS6}
		\epsilon^{(6)} = \left( \mathbf{1} \cos \frac{\psi_1}{2} + i \gamma_1 \sin \frac{\psi_1}{2} \right) \prod_{i=2}^6 \left( \mathbf{1} \cos \frac{\psi_i}{2} + \gamma_{i-1,i} \sin \frac{\psi_i}{2} \right) \epsilon_{\hat{\psi}_1, \cdots, \hat{\psi}_6}^{(6)},
	\end{equation}
	where $\epsilon_{\hat{\psi}_1, \cdots, \hat{\psi}_6}^{(6)}$ doesn't dependent on the coordinates of $S^6$. This motivates us to make the following ansatz for the Killing spinor:
	\begin{equation}
		\begin{aligned}
		\epsilon_{\hat{r}, \alpha} &= \sum_{\sigma \in \mathbb{Z}_2^6 } C_{\sigma, \alpha} f_{\sigma_1} \left( \frac{\psi_1}{2} \right) f_{\sigma_2} \left( \frac{\psi_2}{2} \right) f_{\sigma_3} \left( \frac{\psi_3}{2} \right) f_{\sigma_4} \left( \frac{\psi_4}{2} \right) f_{\sigma_5} \left( \frac{\psi_5}{2} \right) f_{\sigma_6} \left( \frac{\psi_6}{2} \right),\\
		& \qquad f_{\sigma}(x) \equiv \left\{  \begin{array}{c} \cos x,\quad \sigma = 0,\\ \sin x,\quad \sigma = 1. \\  \end{array} \right.
		\end{aligned}
	\end{equation}
	By numerical investigation, we manage to fix all the coefficients $C_{\sigma, \alpha}$ and get the following nice behavior of $\epsilon_{\hat{r}}$:
	\begin{equation} \label{eq:S6dependence}
		\epsilon_{\hat{r}} = \cR_{S^6}(\psi_i, \Gamma) \epsilon_{\hat{r}, \hat{\psi}_1, \cdots, \hat{\psi}_6}, \qquad \cR_{S^6} = \left( \cos \frac{\psi_4}{2} \mathbf{1} - \Gamma^{\textcolor{red}{10, 11}} \sin \frac{\psi_4}{2} \right) \left( \cA \cos \frac{\psi_3}{2} + \cB \sin \frac{\psi_3}{2} \right),
	\end{equation}
	where $\epsilon_{\hat{r}, \hat{\psi}_1, \cdots, \hat{\psi}_6}$ only depends on $\theta$, and the matrices $\cA$ and $\cB$ are as follows: 
	\begin{equation} 
		\begin{aligned}
			\cA &= (c_1 c_6 s_2 s_5-c_2 c_5 s_1 s_6+c_1 c_2 c_5 c_6+s_1 s_2 s_5 s_6)\mathbf{1} + (c_1 c_5 c_6 s_2+c_5 s_1 s_6 s_2-c_1 c_2 c_6 s_5+c_2 s_1 s_5 s_6) \Gamma^{\textcolor{red}{67}}\\
			& + (-c_2 c_5 c_6 s_1-c_6 s_2 s_5 s_1-c_1 c_2 c_5 s_6+c_1 s_2 s_5 s_6)\Gamma^{\textcolor{red}{89}} + (-c_5 c_6 s_1 s_2+c_1 c_5 s_6 s_2+c_2 c_6 s_1 s_5+c_1 c_2 s_5 s_6)\Gamma^{\textcolor{red}{810}};\\
			\cB &= (c_1 c_5 c_6 s_2+c_5 s_1 s_6 s_2+c_1 c_2 c_6 s_5-c_2 s_1 s_5 s_6)\Gamma^{\textcolor{red}{68}} + (c_5 c_6 s_1 s_2-c_1 c_5 s_6 s_2+c_2 c_6 s_1 s_5+c_1 c_2 s_5 s_6) \Gamma^{\textcolor{red}{69}}\\
			& + (c_2 c_5 c_6 s_1-c_6 s_2 s_5 s_1+c_1 c_2 c_5 s_6+c_1 s_2 s_5 s_6) \Gamma^{\textcolor{red}{610}} + (-c_1 c_6 s_2 s_5-c_2 c_5 s_1 s_6+c_1 c_2 c_5 c_6-s_1 s_2 s_5 s_6) \Gamma^{\textcolor{red}{611}};\\
			& \quad c_i \equiv \cos \frac{\psi_i}{2},\quad s_i \equiv \sin \frac{\psi_i}{2}.
		\end{aligned}
	\end{equation}
	To summarise, the Killing spinors preserved by the dWNW background \eqref{eq:dWNWSolution} have the following structure:
	\begin{equation}  \label{eq:KillingSpinordWNW}
		\begin{aligned}
			\epsilon &= e^{r/2} e^{f_0/2} \cR_{S^6}(\psi_i, \Gamma) \cR(\theta, \Gamma) \epsilon_0(\theta),\quad \eta = \Gamma^{\textcolor{red}{23}} \epsilon,\\
		\end{aligned}
	\end{equation}
	with $\cR_{S^6}$ given in \eqref{eq:S6dependence} and $\epsilon_0$ a normalised spinor only depending on $\theta$. We assume that this form applies to the Killing spinors of all $G_2$ invariant backgrounds and use it to obtain the supersymmetric constraints.
	
	\subsection{The BPS equations}
	
\begin{figure}
		\centering \caption{\rm The dependence of the second component of $\epsilon$ on $\theta$ for $\psi_{i = 1,2,\cdots, 6} = 0$ and $r = 0$.}
		\includegraphics[width=6cm]{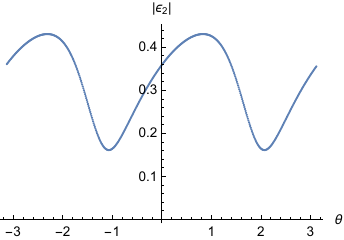}
		\includegraphics[width=6cm]{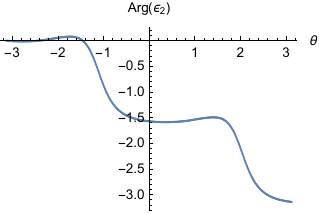}
		\label{fig:epsTheta}
	\end{figure}

	Knowing the dependence on $S^6$, let us choose $\psi_{i = 1,2,\cdots, 6} = 0$ and $r = 0$ in \eqref{eq:KillingSpinordWNW} so that we can focus on the part of the spinor that only depends on $\theta$:
	\begin{equation}
		\epsilon\Big|_{r = \psi_i = 0} = e^{r/2} e^{f_0/2} \cR_{S^6}(\psi_i, \Gamma) \cR(\theta, \Gamma) \epsilon_0(\theta) \Big|_{r = \psi_i = 0} = e^{f_0/2} \cR(\theta, \Gamma) \epsilon_0(\theta).
	\end{equation}
	The dependence on $\psi_{i = 1,2,\cdots, 6}$ and $r$ can be easily restored by \eqref{eq:KillingSpinordWNW}. One component of $\epsilon|_{r = \psi_i = 0}$ as a function of $\theta$ is shown in Fig. \ref{fig:epsTheta}, the analytical form can be obtained by solving the KSE along AdS$_4$ analytically, which is less insightful. We can numerically find the following symmetries of $\epsilon|_{r = \psi_i = 0}$:
	\begin{itemize}
		\item Spinor projectors:
		\begin{equation} \label{eq:SpinorProjectorsRaw}
			\left(\mathbf{1} - \Gamma^{\textcolor{red}{8910\; 11} } \right) \epsilon = 0,\quad 	\left(\mathbf{1} + \Gamma^{\textcolor{red}{67811} } \right) \epsilon  = 0,\quad 	\left(\mathbf{1} - \Gamma^{\textcolor{red}{67910} } \right) \epsilon = 0,
		\end{equation}
		where only two out of them are independent.
		
		\item Charge-conjugation:
		\begin{equation}
			\epsilon =  \Gamma^{\textcolor{red}{156810}} \epsilon^*\quad {\rm or} \quad \epsilon = \Gamma^{\textcolor{red}{23479\,11}} \epsilon^*.
		\end{equation}
		
		\item Inversion symmetry:
		\begin{equation}
			\epsilon = \Gamma^{\textcolor{red}{57810} } \epsilon(-\theta).
		\end{equation}
		The inversion symmetry is more or less expected given the symmetry of the dWNW solution \eqref{eq:dWNWSolution} under $\theta \rightarrow -\theta$. 
	\end{itemize}
	Initially, $\epsilon$ and $\eta$ both has 16 components, then the two spinor projectors above reduce it by a factor of $2^2$, leaving 4 independent free components of $\epsilon$.\footnote{There will be a third projector that reduces the number of components by 2, leaving us only two real independent components. The number of real supercharges is consistent with the 3d $\cN = 1$ superconformal symmetry of the dual field theory. } Thus, the KSE corresponding to each component on $S^6$ \eqref{eq:KSE} is a linear equation with four unknowns. For these equations, the conditions for the existence of non-zero solutions are much easier to obtain compared to a set of equations with 32 unknowns. 
	
	We first look at the KSE along $S^6$, which are of the form $(\partial_{\psi_i} + P_{\psi_i}) \epsilon = 0$. We need to use our knowledge of the $\psi_i$-depdendence to simplify the differential term $\partial_{\psi_i} \epsilon$, so that the differential equation reduces to an algebraic equation. From the study before we know that $\epsilon = \cR_{S^6}(\psi_i, \Gamma) \epsilon_{\hat{\psi}}$, with $\epsilon_{\hat{\psi}}$ independent of $\psi_i$, so all the Killing spinor equations along $S^6$ can be written as homogeneous algebraic equations of four unknowns:
	\begin{equation} \label{eq:AlgebraicEquations}
		\begin{aligned}
			(\partial_{\psi_i} + P_{\psi_i}) \epsilon = \left[ \partial_{\psi_i} \cR_{S^6}(\psi_i, \Gamma) + P_{\psi_i}\cR_{S^6}(\psi_i, \Gamma) \right] \epsilon_{\hat{\psi}}.
		\end{aligned}
	\end{equation}
	This is a key step of our analysis. The existence of non-zero solutions, known as the Cramer's rule, requires the determinant of the coefficient matrix to vanish. In practice, it is easier to solve a subset of the four equations, substitute the solution back to the other equations and require them to vanish. Alternatively, we expect the solutions to the algebraic equations \eqref{eq:AlgebraicEquations} to be independent of $\psi_i$, but the equation itself depends on $\psi_i$, this also gives some conditions consistent with those obtained before. The six KSE along $S^6$ consistently give the following condition:\footnote{As a side note, the derivatives of the emblackening functions come from the spin structure \eqref{eq:susyVariation}. }
	\begin{equation} \label{eq:BPS1}
		e^{-2f_1}(f_0'+2f_2')^2 - 4 g_3^2 e^{-8f_2} - 4 e^{-2f_2} + e^{-2f_0} = 0.
	\end{equation}
	The KSE $\delta \Psi_t = 0$, which can be simplified into \eqref{eq:AlgebraicKSEFull}, is already an algebraic equation. Similar steps impose the following condition:
	\begin{equation}
		4 e^{-2f_1} f_0'f_2' + e^{-2f_1}f_0'^2 \left( 3 + g_3^{-2} e^{6f_2} \right) + \frac{1}{9} g_0^2 e^{-8f_0} + 4 g_3^2 e^{-8f_2} - e^{-2f_0} = 0,
	\end{equation}
	where we have used the first condition \eqref{eq:BPS1} to simplify it. But we haven't finished. We expect the two sets of equations, $\delta \Psi_{\psi_i} = 0$ and $\delta \Psi_t = 0$, to admit the same set of solutions. Requiring the consistency between their solutions, we get the following constraint:
	\begin{equation}
		2g_3g_3' + 12 g_3^2 f_2' + \left( 12g_3^2 + 3e^{6f_2} \right)f_0' = 0.
	\end{equation}
	Since we don't know the dependence of $\epsilon$ on $\theta$ analytically, we can't extract any useful information from the KSE $\delta \Psi_\theta = 0$ as we can't further simplify its differential term $\partial_\theta \epsilon$. But luckily enough, the three equations above already imply the equations of motion \eqref{eq:eomComponents}, as can be checked directly, and thus comprise the BPS conditions we are after. In fact, the BPS equations are not completely equivalent to the equations of motion: the BPS equations imply the equations of motion but not vice versa, meaning that we may miss some non-supersymmetric solutions of the equations of motion which do not satisfy the BPS equations. The non-supersymmetric solutions with SO(7) isometry \cite{Englert:1982vs, deWit:1984va, Duboeuf:2024tbd} are the examples. But we expect the BPS equations to admit all the supersymmetric saddles, such as the AdS$_4 \times S^7$ and the dWNW background.
	
	Out of the unknown functions $f_{0,1,2}, g_{1,2,3}$ in our ansatz, $f_1$ is totally free up to redefinition of $\theta$, and $g_1, g_2$ only appear in the combination $g_2' - 3g_1$ and thus correspond to one freedom. So we have four functions and four equations:
	\begin{equation} \label{eq:BPSEquations}
		\begin{aligned}
				& 4 e^{-2f_1} f_0'f_2' + e^{-2f_1}f_0'^2 \left( 3 + g_3^{-2} e^{6f_2} \right) + \frac{1}{9} g_0^2 e^{-8f_0} + 4 g_3^2 e^{-8f_2} - e^{-2f_0} = 0, \\
				& e^{-2f_1}(f_0'+2f_2')^2 - 4 g_3^2 e^{-8f_2} - 4 e^{-2f_2} + e^{-2f_0} = 0, \\
				& 	2g_3g_3' + 12 g_3^2 f_2' + \left( 12g_3^2 + 3e^{6f_2} \right)f_0' = 0, \\
				&  \left( g_2' - 3g_1\right) = e^{f_1-4f_0} g_0 g_3, \\
		\end{aligned}
	\end{equation}
	which is completely solvable in principle. Since the last equation is an algebraic one, in the next section, we will focus on the first three equations and try to solve them analytically by power expansion around special values of $\theta$ and then numerically. This will reproduce the known SO(8) and $G_2$-invariant solutions \eqref{eq:AdS4S7} \eqref{eq:dWNWSolution} and generate a new $G_2$ invariant solution.

\section{The $G_2$-invariant solutions}
\label{sec:G2Saddles}

	In this section, we show that the BPS equations give the full set of $G_2$-invariant saddles in 11 dimensional supergravity that preserve at least four supercharges. They include the maximally supersymmetric SO(8)-invariant saddle and two $G_2$ saddles that preserve four supercharges.
	
	We start with the Freund-Rubin case, where $g_3 = 0$. The BPS equations \eqref{eq:BPSEquations} immediately dictate $f_0$ to be a constant, reducing the problem to the special case already discussed in \cite{Gunaydin:1983mi} and no new solution is found. The SO(8)-invariant saddle is reproduced in this special case. For the other saddles, we expect the internal flux to be turned on, i.e., $g_3$ to be a non-trivial function. 
	
	In the following section \ref{StationaryLocus}, we highlight some properties of the first three BPS equations as two dynamical systems and motivate the subsequent study on the integral curves. In section \ref{subsec:theta=0}, we start with the family of integral curves that caps off smoothly at $\theta = 0$ and reproduce the dWNW saddle. Then in section \ref{subsec:IntegralCurvesBranch2}, we introduce another family of integral curves that stems from a stationary point of the dynamical system. A numerical shooting exercise gives the new $G_2$ saddle.  In Appendix \ref{App:DynamicalSystem}, we will give a detailed analysis which shows the absence of any other solutions with expected properties and presents a new $G_2$-invariant solution whose internal space is squashed $S^1 \times S^6$ periodic in $\theta$.
	
	\subsection{The BPS equations as two dynamical systems}
	
	We notice that the first two BPS equations \eqref{eq:BPSEquations} can be regarded as quadratic algebraic equations of $f_0'$ and $f_2'$ with only quadratic and constant terms, so they are invariant under $(f_0', f_2') \rightarrow - (f_0', f_2')$. The third equation only involves $\tilde{g}_3 \equiv g_3^2$ and its derivative and is thus invariant under $g_3(\theta) \rightarrow - g_3(\theta)$, so we can assume $g_3 \ge 0$ without loss of generality. Up to the symmetries, the quadratic equations of $f_0'$ and $f_2'$ give two sets of solutions, both of which are of the form of a dynamical system with variables $\left( e^{f_0}, e^{f_2}, g_3\right)$:
	\begin{equation} \label{eq:DynamicalSystem}
		f_0' = \cF_0^{(1),(2)} \left( e^{f_0}, e^{f_2}, g_3\right),\quad f_2' = \cF_2^{(1),(2)} \left( e^{f_0}, e^{f_2}, g_3\right),\quad g_3' = \cG_3^{(1),(2)} \left( e^{f_0}, e^{f_2}, g_3\right),
	\end{equation}
	where $\cF_0^{(1),(2)}, \cF_2^{(1),(2)}, \cG_3^{(1),(2)}$ are some explicit algebraic expressions of $ e^{f_0}, e^{f_2}$, and $g_3$ with square roots. We will call the two dynamical systems \DS1 and \DS2 in what follows. \DS1 defined as the one that contains the dWNW solution.
	
	As the first two BPS equations are quadratic equations of $f_0'$ and $f_2'$, the existence of real solutions imposes the following two constraints:
	\begin{equation} \label{eq:RealityConditions}
		4 e^{2f_0} (e^{6f_2} + g_3^2 ) \ge e^{8f_2},\quad (9e^{6f_0} - 1)e^{6f_2} \ge g_3^2.
	\end{equation}
	The locus where at least one of the conditions is saturated form codimensional-one surfaces shown in the left panel of Fig. \ref{fig:IntegrationCurves} and the Dynamical Systems are real if and only if bounded by the two surfaces. The two inequalities are saturated on the green and orange surfaces respectively. On the surfaces, the quadratic equations of $f_0'$ and $f_2'$ are degenerate, meaning that the two Dynamical Systems are identical on the boundary surfaces. 
	
	The internal manifold of the geometry \eqref{eq:ansatzMetric} is a warped product of $S^6$ and an interval $I_\theta$. As we expect the internal manifold to have the topology of a sphere, $f_2$ should vanish at both ends of the interval $I_\theta$. The smoothness of $f_2$ then suggests an extremal point $\theta_* \in I_\theta$ where $f_2'(\theta_*) = 0$. Thus, we may ask for what values of $(e^{f_0}, e^{f_2}, g_3)$ do we have $f_2' = \cF_2^{(1), (2)} (e^{f_0}, e^{f_2}, g_3) = 0$ in the two dynamical systems.
	
	{\bf \DS 1}
	
	For \DS 1, requiring $f_2'(e^{f_0}, e^{f_2}, g_3) = 0$ gives us two branches of one-dimensional families (i.e., co-dimensional 2 curves in the domain of the dynamical system) of solutions parametrized by $\alpha \equiv e^{f_0}(\theta_*)$:
	\begin{itemize} \label{StationaryLocus}
		\item Branch 1: 
		\begin{equation}
		g_3(\theta_*) = (6 \alpha^4)^3 \sqrt{9 \alpha^6 - 1}, \quad \beta = 6 \alpha^4,\quad \beta \equiv e^{f_2}(\theta_*).
		\end{equation}
		
		\item Branch 2: 
		\begin{equation} \label{eq:branch2}
		g_3(\theta_*) =0, \quad  \beta= 2 \alpha.
		\end{equation}
	\end{itemize}
	In fact, on the branches we have not only $f_2' = 0$ but also $f_0' = \tilde{g}_3' = 0$, i.e., the points on the branches are stationary points of the dynamical system. Pictorially, Branch 1 represents the red curve in Fig. \ref{fig:IntegrationCurves} which is the intersection of the green and orange surfaces, and Branch 2 represents the red straight line in Fig. \ref{fig:IntegrationCurves} which is the intersection of the green surface and the $g_3 = 0$ plane. Taking a point on Branch 1 as the initial value of the dynamical system, we get three families of solutions, two of which are on the degenerate surfaces and the third one contains the dWNW solution and AdS$_4\times S^7$.  Starting from a point on Branch 2, there are also three families of solutions, two of which lie on the two degenerate surfaces respectively and the other one actually belongs to \DS 2 which gives a new $G_2$ invariant solution. We will study the two branches with full details in section \ref{subsec:f2saddle}.
	
	{\bf \DS 2}
	
	For \DS 2, the condition $f_2'(\theta_*) = 0$ is difficult to reduce. By power expansion over $(\theta - \theta_*)$ and focusing on the leading order, we obtain a two-parameter family of solutions, whose parameters can be chosen as:
	\begin{equation*}
		\alpha \equiv e^{f_0}(\theta_*),\quad \beta \equiv e^{f_2}(\theta_*),
	\end{equation*}
	and the surface where $f_2' = 0$ is given by:
	\begin{equation} \label{eq:Solution2}
	 g_3(\theta_*)^2 = \frac{\beta^6}{288\alpha^8} \left[ (36\alpha^6 - 1) \beta^2 - 144\alpha^8 + \beta \sqrt{(36\alpha^6-1)^2\beta^2 - 288\alpha^8 (18\alpha^6-1) }\right].
	\end{equation}
	Note that we don't have $f_0' = g_3' = 0$ at these points. For a chosen value of $\alpha \ge \frac{1}{3^{1/3}}$, the square root on the right-hand side dictates the range of $\beta$ to be $2\alpha \le \beta \le 6\alpha^4$. Notice that the upper and lower bound of $\beta$ respectively correspond to Branch 1 and 2 above, on which the two dynamical systems are identical. For any initial value $(\alpha, \beta)$ within the domain, one can numerically integrate the BPS equations towards two directions. A typical integral curve of \DS2 connects the green and the orange surfaces, as shown in Fig. \ref{fig:IntersectionS1S2}.
	
	\subsection{Integral curves starting from $\theta = 0$ and the known saddles}
	\label{subsec:theta=0}
	
	The range of $\theta$ begins at $\theta = 0$ where $f_2(\theta)$ goes to zero linearly. The most general small-$\theta$ expansion is:
		\begin{equation}
		e^{f_0} = \sum_{i=0}^\infty \tilde{f}_{0,i} \theta^i,\quad e^{f_1} = \sum_{i=0}^\infty \tilde{f}_{1,i} \theta^i,\quad e^{f_2} = \sum_{i=1}^\infty \tilde{f}_{2,i} \theta^i,\quad \tilde{g}_3 \equiv g_3^2 =  \sum_{i=0}^\infty \tilde{g}_{3,i} \theta^i.
	\end{equation}
	Since the BPS equations \eqref{eq:BPSEquations} only involve $\tilde{g}_3 = g_3^2$ and its derivative, we find it more convenient to deal with $\tilde{g}_3$ directly. We need to fix $f_1$ to make further evaluation, which amounts to fixing the reparametrisation of the $\theta$-coordinate. For the two known solutions, $f_1$ is related to $f_0$ in different ways:
	\begin{equation} \label{eq:f1Convention}
		{\rm SO(8)}:\quad e^{f_1} = 2 e^{f_0},\qquad G_2: \quad e^{f_1} = \frac{2^{3/2}}{5^{1/2}} e^{f_0}.
	\end{equation}
	We will follow the convention of the dWNW $G_2$ solution, which amounts to setting $\tilde{f}_{1,i} =  \frac{2^{3/2}}{5^{1/2}} \tilde{f}_{0,i}$ for all $i = 0,1,\cdots$ in the expansion around $\theta = 0$. By solving the BPS equations order by order in terms of $\theta$-expansion, we get one family of solutions parametrized by $\tilde{f}_{0,0}$, the first few orders are given by:
	\begin{equation} \label{eq:PowerExpansionFamily4}
		\begin{aligned}
			e^{f_0} &= \tilde{f}_{0,0} - \frac{4(9\tilde{f}_{0,0}^6 - 1)}{3\cdot 7^2 \tilde{f}_{0,0}^5 } \theta^2 + \frac{8  \left(9 \tilde{f}_{0,0}^6-1\right)\left(65 \tilde{f}_{0,0}^6 + 69 \right) \theta^4 }{324135\tilde{f}_{0,0}^{11}} + O(\theta)^6, \\
		e^{f_2} &= \frac{2^{3/2}}{5^{1/2}} \, \theta \left[   \tilde{f}_{0,0} -\frac{\left(40  \tilde{f}_{0,0}^6+1\right) \theta^2 }{735  \tilde{f}_{0,0}^5} + \frac{\left(201-1760 \tilde{f}_{0,0}^6\right)  \theta^4 }{330750 \tilde{f}_{0,0}^{11}} + O( \theta)^6 \right],\\
		\tilde g_3 \equiv g_3^2 &= \theta^8 \left[ \frac{1024 \left(9 \tilde{f}_{0,0}^6-1\right)}{5^4 \cdot 7^2} + \frac{65536 \left(9 \tilde{f}_{0,0}^6-1\right)\left(5 \tilde{f}_{0,0}^6-6 \right) \theta^2 }{67528125
	\tilde{f}_{0,0}^6} + O( \theta)^4 \right].
		\end{aligned}
	\end{equation}
	Notice that the expansions are given in $\theta^2$ instead of $\theta$, reflecting the symmetry $(f_0', f_2') \rightarrow - (f_0', f_2')$. We expect $\tilde{g}_3  = g_3^2 \ge 0$, whose leading term in $\theta$-expansion dictates $\tilde{f}_{0,0} \ge \frac{1}{3^{1/3}} \approx 0.693$. Taking the limit $\tilde{f}_{0,0} \rightarrow \frac{1}{3^{1/3}}$, we find that all $\tilde{g}_{3,i}$, and thus $\tilde{g}_3$, vanish: the solution boils down to the ${\rm AdS}_4 \times S^7$ solution \eqref{eq:AdS4S7}. \footnote{\label{footnote:rangeofTheta}A subtlety here: since we take $e^{f_1} = \frac{2^{3/2}}{5^{1/2}} e^{f_0}$ during the numerics, which is inconsistent with the AdS$_4 \times S^7$ saddle shown in \eqref{eq:AdS4S7}, we need to rewrite the solution as:
		\begin{equation} \label{eq:AdS4S7Theta}
			e^{f_0}  =  \frac{1}{3^{1/3}}, \quad e^{f_1} = \frac{2^{3/2}}{5^{1/2}} e^{f_0}, \quad e^{f_2} = \frac{2}{3^{1/3}} \sin\left( \sqrt{\frac{2}{5}} \theta \right).
		\end{equation} } The $G_2$ invariant dWNW solution \eqref{eq:dWNWSolution} corresponds to $\tilde{f}_{0,0} = \left(\frac{6}{5} \right)^{1/6} \approx 1.031$, when the second term in the $\theta$-expansion of $\tilde g_3$ vanishes, interestingly.
	
	\begin{figure}[!h]
		\centering \caption{\rm We showcase the integral curves starting at $\theta = 0$ and $e^{f_2} = 0$. The coordinates are values of $e^{f_0}$, $e^{f_2}$ and $g_3$. The red curves are locus where we have $f_2' = 0$ found in section \ref{StationaryLocus}. The blue curve and blue line indicate the dWNW and AdS$_4\times S^7$ solutions. Left panel: The green and orange surfaces illustrate locus where the BPS equations for $f_0'$ and $f_2'$ \eqref{eq:DynamicalSystem} are degenerate and the inequalities \eqref{eq:RealityConditions} are saturated. The family of curves in orange indicates integration curves with different initial values $\tilde{f}_{0,0}$. Right panel: We zoom in to see the positions of dWNW and AdS$_4\times S^7$ solutions.  }
		\includegraphics[width=6cm]{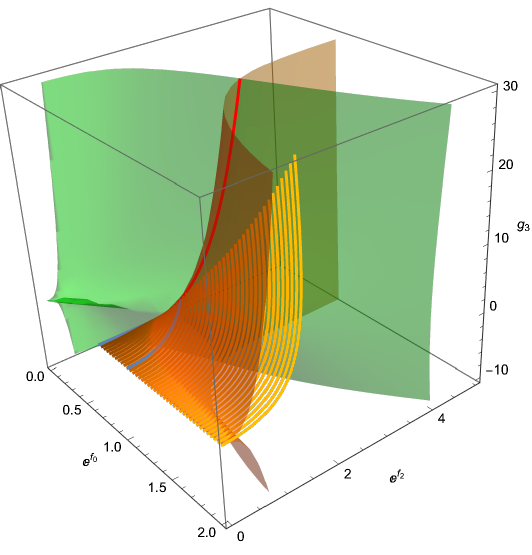} \hspace{0.5cm}
		\includegraphics[width=6cm]{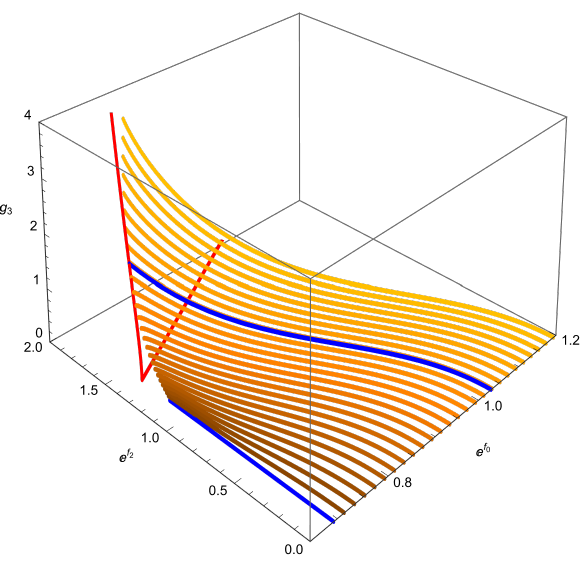}
		\label{fig:IntegrationCurves}
	\end{figure}
	
	We did not assume the scalar curvature to be finite everywhere, but we notice that this condition is already dictated by the BPS conditions. The scalar curvature for the metric ansatz \eqref{eq:ansatzMetric} is:
	\begin{equation*}
		R = -2e^{-2f_1}\left[4 f_0'' + 6 f_2'' + 10(f_0')^2 + 21 (f_2')^2 - 4f_0'f_1' + 24 f_0'f_2' - 6f_1'f_2' \right] - 12 e^{-2f_0} + 30 e^{-2f_2}.
	\end{equation*}
	Plugging in the small-$\theta$ expansion and requiring the scalar curvature to be finite at $\theta \rightarrow 0$, we get the following conditions:
	\begin{equation}
		\tilde{f}_{2,1} = \tilde{f}_{1,0},\quad \tilde{f}_{2,2} = \frac{1}{2}\tilde{f}_{1,1} - \frac{\tilde{f}_{0,1} \tilde{f}_{1,0} }{3 \tilde{f}_{0,0} } ,
	\end{equation}
	which are automatically satisfied by the expansion \eqref{eq:PowerExpansionFamily4}. 
	
	Taking the first few orders in the power series expansion as the initial conditions, we can numerically integrate the BPS conditions and get integral curves in the space $(e^{f_0}, e^{f_2}, g_3)$, see Fig. \ref{fig:IntegrationCurves} for an illustration. It's easy to see that the AdS$_4 \times S^7$ and dWNW solutions belong to this family of integral curves and correspond to $\tilde{f}_{0,0} = \frac{1}{3^{1/3}}$ and $\left( \frac{6}{5} \right)^{1/6}$, respectively. The initial value corresponding to the AdS$_4 \times S^7$ solution has the minimal $\tilde{f}_{0,0}$. The dWNW solution plays an important role in this family: for an integral curve with larger $\tilde{f}_{0,0}$, it reaches the green surface where the numerical algorithm breaks down\footnote{This doesn't mean the integral curve terminates on the green surface. In fact, as we will show in section \ref{sec:ConnectingFamilies}, the integral curve continues flowing on the boundary surface until it reaches a stationary point. However, the curve doesn't smoothly transits to the surface. }; while for a curve with a smaller $\tilde{f}_{0,0}$, it reaches the orange surface; the dWNW solution itself reaches the intersection of the two surfaces. 
	
	As argued before, to have a solution whose internal space has spherical topology, we expect that there is a special value $\theta = \theta_*$ where $f_2'(\theta_*) = 0$. Since these integral curves are of \DS 1, this is only possible if the integral curve reaches one of the branches we find in section \ref{StationaryLocus}, which corresponds to the red curves in Fig. \ref{fig:IntegrationCurves}. Apparently, the only integration curve satisfying this corresponds to the dWNW solution; all the other integration curves interset with one of the two surfaces where the \DS is degenerate. To find new solutions, we need to study integral curves of \DS 2 as well. 
	
	\begin{figure}[!h]
		\centering \caption{\rm {\it Left panel:} The AdS$_4 \times S^7$ solution \eqref{eq:AdS4S7Theta} plotted in terms of functions of $\theta \in (0, \sqrt{\frac{5}{2}}\pi)$. {\it Right panel:} The dWNW solution \eqref{eq:dWNWSolution} plotted in terms of functions of $\theta \in (0, \pi)$. The free parameter $g_0$ is set to 1. }
		\includegraphics[width=7cm]{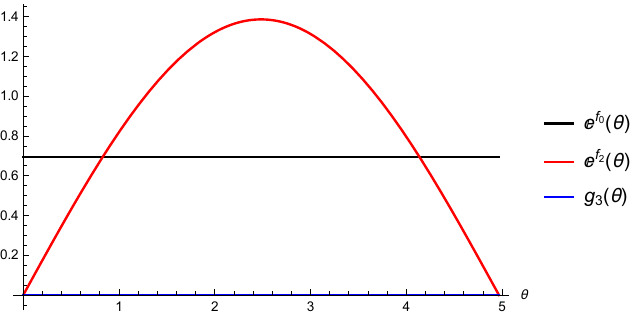}  \hspace{0.5cm}
		\includegraphics[width=7cm]{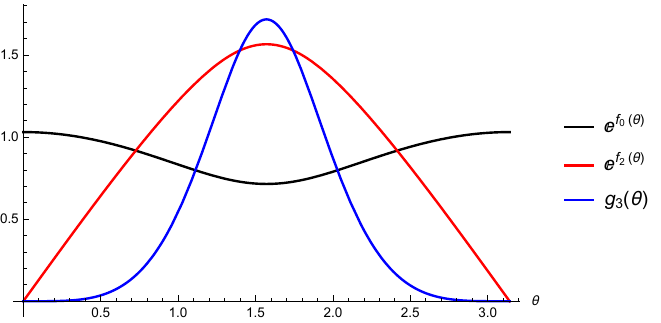} 
		\label{fig:dWNWTheta}
	\end{figure}
	
	\subsection{Integral curves starting from Branch 2 and the new $G_2$-saddle} 
	\label{subsec:IntegralCurvesBranch2}
	
	We consider the most general ansatz for the three functions expanded around $\theta = \theta_*$ where $f_2'(\theta_*) = 0$: \footnote{Although we use the same notation $\hat{f}_{0,0}$ for different families of solutions around $\theta_*$, they should be considered as independent. }
	\begin{equation*}
		e^{f_0} = \sum_{i=0}^\infty \hat{f}_{0,i} (\theta - \theta_*)^i,\quad e^{f_1} = \frac{2^{3/2}}{5^{1/2}} e^{f_0},\quad e^{f_2} = \sum_{i=0, i\ne 1}^\infty \hat{f}_{2,i} (\theta - \theta_*)^i,\quad \tilde g_3 = \sum_{i=0}^\infty \hat{g}_{3,i} (\theta - \theta_*)^i.
	\end{equation*}
	The BPS equations in the leading order power expansion gives the two options:
	\begin{equation}
		\hat{f}_{2,0} = 6 \hat{f}_{0,0}^4,\quad {\rm or} \quad \hat{f}_{2,0} = 2 \hat{f}_{0,0},
	\end{equation}
	which correspond exactly to Branch 1 and Branch 2 we found in section \ref{StationaryLocus}. Starting from a point of Branch 1 or Branch 2, we find in total five families of integral curves. We will focus on a specific family of solutions starting from Branch 2, and refer the other families to Appendix \ref{App:DynamicalSystem}. The specific family of solutions has the following power series expansion:
	\begin{equation} \label{eq:ExpansionFamily3Branch2}
		\begin{aligned}
			e^{f_0} &= \hat{f}_{0,0} - \frac{2\left( 9 \hat{f}_{0,0}^6 - 1 \right)\rho }{15 \hat{f}_{0,0}^{5} } + \frac{2 \left( 9 \hat{f}_{0,0}^6 - 1 \right)\left( 39 \hat{f}_{0,0}^6 + 2 \right) \rho^2 }{ 225 \hat{f}_{0,0}^{11} } + O(\rho)^3,\quad \rho \equiv (\theta - \theta_*)^2,\\
			e^{f_2} &= 2 \hat{f}_{0,0} +  \frac{2 \left( 6 \hat{f}_{0,0}^6 - 1 \right)\rho }{15 \hat{f}_{0,0}^5 } - \frac{  \left( 996  \hat{f}_{0,0}^{12} -48  \hat{f}_{0,0}^{6} -7   \right) \rho^2 }{225 \hat{f}_{0,0}^{11} } + O(\rho)^3, \\
			\tilde g_3 &= \frac{128}{5} \left( 9 \hat{f}_{0,0}^6 - 1 \right) \left[ \rho + \frac{4 \left( 5 \hat{f}_{0,0}^6 - 1 \right) \rho^2 }{5 \hat{f}_{0,0}^{6} }  + \frac{4 \left( 152 \hat{f}_{0,0}^{12} - 137 \hat{f}_{0,0}^{6} + 15 \right) \rho^3 }{125 \hat{f}_{0,0}^{12}} + O(\rho)^4 \right].
		\end{aligned}
	\end{equation}
	There is one free parameter $\hat{f}_{0,0} \equiv e^{f_0}(\theta_*)$ that parametrizes this family. With the above initial condition, we can numerically integrate the dynamical system \eqref{eq:DynamicalSystem}. One complication is that this family of integral curves belongs to \DS2, where there is a special surface across which the integral lines change direction and thus be non-analytic:
	\begin{equation} \label{eq:NonAnalyticSurface}
		\begin{aligned}
			g_3 = \frac{1}{6} e^{-4f_0 + 4f_2} \sqrt{9 e^{6f_0} - 1}.
		\end{aligned}
	\end{equation}
	\begin{figure}[h]
		\centering \caption{\rm We showcase the integral curves of the third family of solutions belong to Branch 2, with the series expansion in \eqref{eq:ExpansionFamily3Branch2}. The vertical coordinate is $g_3$, and the left and right panels are from the perspective of $e^{f_0}$-axis and $e^{f_2}$-axis, respectively. The green and orange surfaces are defined as before. The family of curves in pink indicates integration curves with different initial values $\hat{f}_{0,0}$. }
		\includegraphics[width=7cm]{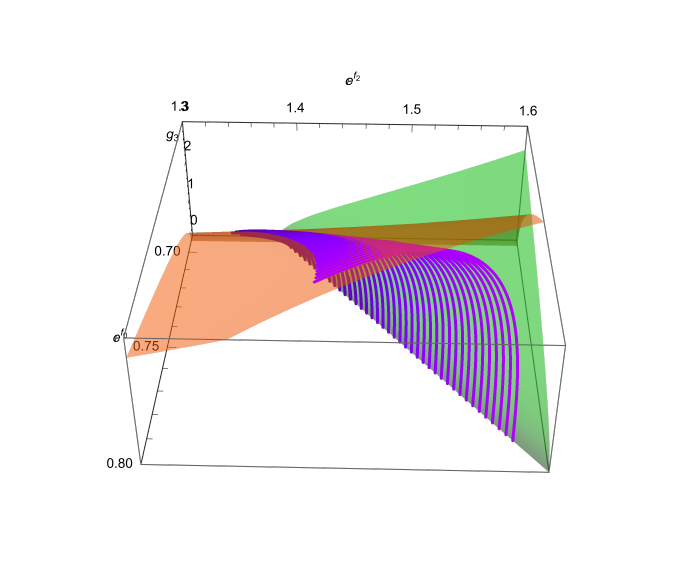}
		\includegraphics[width=7cm]{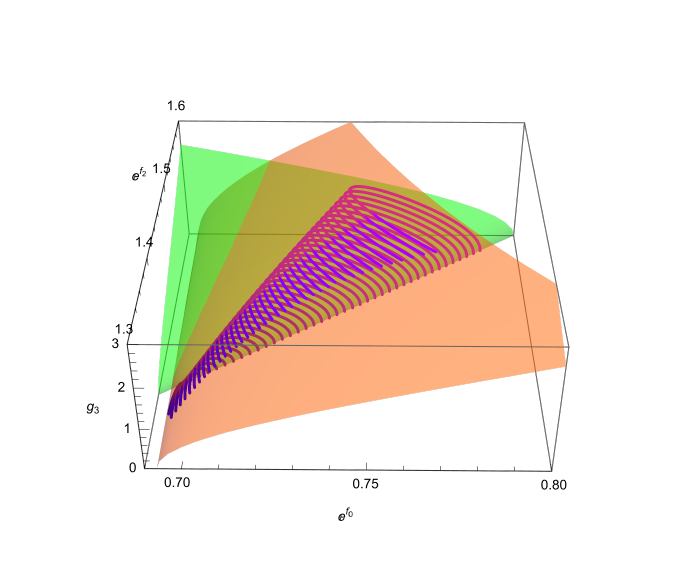}
		\label{fig:Branch2Family3}
	\end{figure}
	In our numerical manipulation for the integral curves, we need to move the integration curve across the surface by hand. The final numerical results are presented in Fig. \ref{fig:Branch2Family3}, where we can see that all the curves finally reach the orange surface. The behavior of the dynamical system on the orange surface will be studied in detail in Appendix \ref{App:DynamicalSystem}, with the conclusion that the family of integral curves on their own don't generate continuous solutions.\footnote{A special case is the limit where $\hat{f}_{0,0}$ approaches the smallest allowed value $\frac{1}{3^{1/3}}$, which gives the AdS$_4 \times S^7$ solution. }
	
		\begin{figure}[!h]
		\centering 
		\caption{\rm The orange-yellow family of integral curves starts from $\theta = 0$ with initial conditions \eqref{eq:PowerExpansionFamily4}, and the violet-pink family of integral curves start from Branch 2 with initial conditions \eqref{eq:ExpansionFamily3Branch2}.  } 
		\includegraphics[width=7cm]{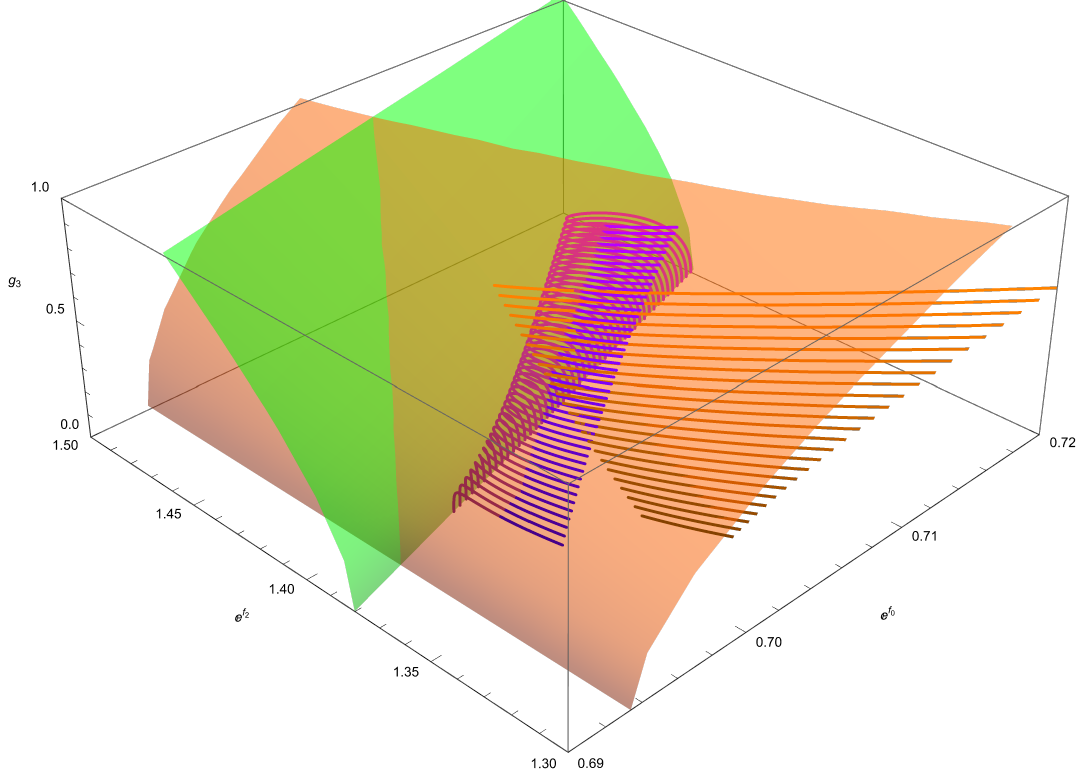}
		\label{fig:ConnectingFamilies}
	\end{figure}
	
\begin{figure} [!h]
		\centering 
		\caption{\rm We show the continuous solution with initial value $\hat{f}_{0,0} \approx 0.879040 ,\quad	\hat{f}_{0,0} \approx 0.703761$ obtained from the numerical shooting process. } 
		\includegraphics[width=4.5cm]{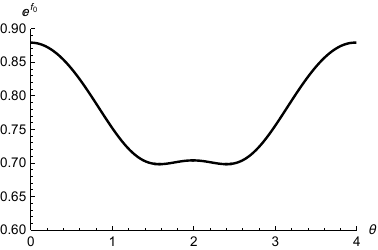} \hspace{0.3cm}
		\includegraphics[width=4.5cm]{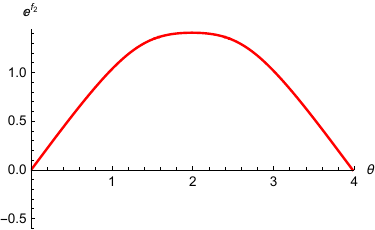} \hspace{0.3cm}
		\includegraphics[width=4.5cm]{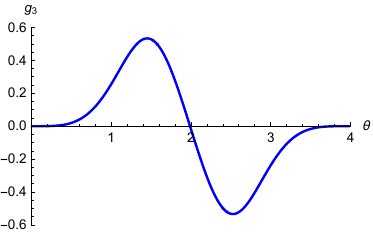}
		\label{fig:G2p}
	\end{figure}

	It is natural now to consider the combination of this family of integral curves, which start from an extremal point of $f_2(\theta)$, with the family of integral curves discussed in section \ref{subsec:theta=0}, which start from $\theta = 0$. Since this family of integral curves all flow to the orange surface, we only need to consider initial values $\tilde{f}_{0,0} < \tilde{f}_{0,0}^{\rm dWNW}$ in the family of section \ref{subsec:theta=0}. As shown in Fig. \ref{fig:ConnectingFamilies}, there are infinitely many pairs of integral curves from the two families which intersect, but since they belong to two different Dynamical Systems, they are not smoothly connected unless the Dynamical Systems are degenerate, which only happens on the green and orange surfaces. As can be seen from the plot, there is only one pair of integral curves that meet on the orange surface. Numerically, we identify the corresponding initial values:
	\begin{equation}
		\tilde{f}_{0,0} = 0.879040\cdots ,\quad	\hat{f}_{0,0} = 0.703761 \cdots.
	\end{equation}
	The corresponding new $G_2$-invariant solution is shown in Fig. \ref{fig:G2p}. It is also important to make sure that the functions $f_0, f_2, g_3$ are smooth everywhere, i.e., both the first-order and the second-order derivatives in terms of $\theta$ are continuous, especially on the two sides of the intersection point. We have indeed verified this numerically, making sure that it is a smooth solution. Since the internal space is a squashed seven-sphere preserving $G_2$ isometry, it is a strong candidate of the IR fixed point of the holographic RG flow that motivates our investigation. We will analyse the holographic properties of the new saddle in the next section.
	
\section{The web of holographic RG flows}
\label{sec:freeEnergy}

	Since we have identified a candidate of the gravity dual of the $G_2$-invariant IR fixed point, it is interesting to study what predictions they give to the field theory. Here we study the holographic free energy in both some of the known supergravity saddles dual to RG fixed points and our new solution, completing the web of RG flows shown in Fig. \ref{fig:G2FixPt}.
	
	\subsection{SU(3) $\times$ U(1) fixed points}
	On the field theory side, the supersymmetric free energy of the $\cN = 2$ massive ABJM theory deformed by the single-trace operator ${\rm tr} X^p$ has been evaluated by supersymmetric localisation. The ratio of the UV and IR free energies should be: \cite{Jafferis:2011zi}
	\begin{equation}
		\frac{F^{\rm IR}}{F^{\rm UV}} = \frac{16(p-1)^{3/2}}{3\sqrt{3}p^2 } = \left\{ \begin{aligned}
		& \frac{2^{2}}{3^{3/2}} = 0.7698\cdots ,\quad p = 2;\\
		& \frac{2^{11/2}}{3^{7/2}} = 0.9677\cdots ,\quad p = 3. \\
		\end{aligned}\right.
	\end{equation}
	
\begin{figure}
		\centering 
		\caption{\rm Left: the CPW solution \cite{Corrado:2001nv}, Right: the GMPS solution \cite{Gabella:2012rc}, where the vertical line comes out of the limited numerical precision and should vanish in the exact solution.  } 
		\includegraphics[width=7cm]{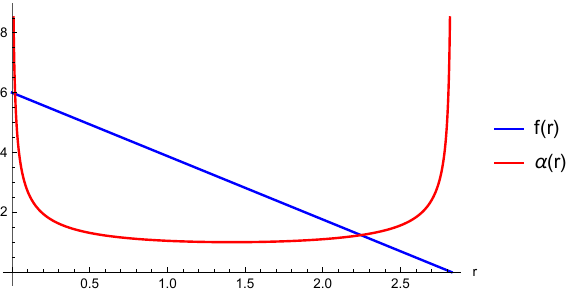} \hspace{0.5cm}
		\includegraphics[width=7cm]{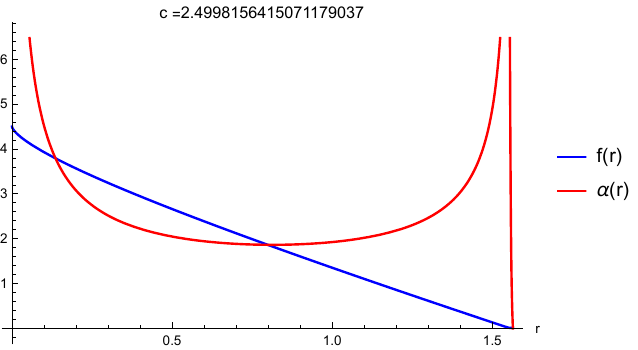}
		\label{GMPSandCPW}
	\end{figure}

	The 11 dimensional supergravity dual to the fixed points are the CPW \cite{Corrado:2001nv} and GMPS \cite{Gabella:2012rc, Halmagyi:2012ic} solutions for $p = 2$ and $3$, respectively. Following the notation of  \cite{Gabella:2012rc}, the metric and flux are:
	\begin{equation}
		\begin{aligned}
			ds^2_{11} &= \frac{1}{4} e^{2\Delta} ds_4^2 + e^{2\Delta} ds_7^2, \\
			ds^2_7 &= \frac{f\alpha}{4\sqrt{1+(1+r^2)\alpha^2}} ds^2_{\mathbb{CP}^2} + \frac{\alpha^2}{16} \left[ dr^2 + \frac{r^2f^2}{1+r^2} (d\tau + \cA)^2 \right.\\
			& \left.\hspace{5cm}  + \frac{1+r^2}{ 1+(1+r^2)\alpha^2 } \left( d\psi + \frac{f}{1+r^2} (d\tau + \cA) \right)^2 \right],\\
			e^{6\Delta} &= \left( \frac{m}{6} \right)^2 \frac{ 1+ (1+r^2)\alpha^2 }{\alpha^2},\quad  0 < \tau < 2\pi,\quad 0 < \psi < 4\pi, \\
	 F &= \frac{m}{16} {\rm vol}_4 + F_{\rm internal},
		\end{aligned}
	\end{equation}
	where $ds^2_{\mathbb{CP}^2}$ and $\cA$ are the standard Fubini-Study metric and Kaehler form on $\mathbb{CP}^2$. The planar holographic free energy is given by: \cite{Gabella:2012rc}
	\begin{equation} \label{eq:Fleading}
		\begin{aligned}
			F &= \sqrt{\frac{m^3\pi^6}{2^23^6\int e^{9\Delta}{\rm vol}_7}} N^{3/2} =  \frac{2^{\frac{9}{2}} \pi }{3^{\frac{3}{2}} \sqrt{ \int f^3 \alpha^2 r dr }} N^{3/2}, \\
		\end{aligned}
	\end{equation}
	where $N$ is the quantized M2-brane charge. The CPW solution is known analytically:
	\begin{equation}
		\alpha(r) = \sqrt{\frac{2}{r(2\sqrt{2} - r) }},\quad f(r) = \frac{3}{\sqrt{2}}  \left( 2\sqrt{2} - r \right),\quad r\in [0, 2\sqrt{2}].
	\end{equation}
	Then it's straightforward to evaluate that\footnote{The holographic free energy for AdS$_4\times S^7$ is given by \cite{Herzog:2010hf}:
	\begin{equation}
		F_{\rm SO(8)} = \frac{\sqrt{2}\pi }{3}N^{3/2}.
		\end{equation} }
	\begin{equation}
		 F_{\rm CPW} = \left(\frac{2}{3}\right)^{5/2} \pi N^{3/2}\quad \Rightarrow \quad \frac{F_{SU(3)\times U(1)}}{F_{SO(8)}} = \frac{2^2}{3^{\frac{3}{2}} } = 0.7698\cdots ,
	\end{equation}
	identical to the field theory result. For the GMPS solution, we need to solve the BPS equations numerically, using the initial condition given in \cite{Gabella:2012rc} \footnote{ There's a typo in the last term of equation (4.63) there. } with $\rho \equiv r^{2/3}$ as a new variable. By requiring $f(r)$ to be linear at $r = r_0$ on the right-hand side of the domain, we perform numerical shooting and get $c \approx 2.4998156 \cdots$, where the solution is shown in Fig. \ref{GMPSandCPW}. 
	
	With the numerical solution at hand, we can evaluate the holographic free energy numerically:\footnote{
		Note that the range of $(\psi, \tau)$ is different for different values of $p$, as discussed in section 2.2 of \cite{Cesaro:2020piw}. Roughly speaking, the range of $\psi$ is $2\pi p$ instead of fixed to $4\pi$. What's more, in the same paper, the numerical plot of $\alpha(R)$ seems to be wrong. 
	}
	\begin{equation}
		\frac{F^{p = 3}_{SU(3)\times U(1) }}{N^{3/2}} = 1.433135 \cdots ,\quad \Rightarrow \quad \frac{F^{p=3}_{SU(3)\times U(1)}}{F_{SO(8)}} =  0.9677 \cdots,
	\end{equation}
	which also reproduces the field theory result.
	
	    \subsection{The $G_2$-invariant solutions }
	 For the web of RG flow to be valid, the $F$-theorem \cite{Jafferis:2011zi} requires that the free energy of the $G_2$-invariant solution lies between the UV and the IR fixed points:
	 \begin{equation}
	 	F_{\rm SO(8)} > F_{\rm G_2} > F_{\rm SU(3)\times U(1)}.
	 \end{equation}
	Using Table \ref{tbl:ComparisonConventions} to match the conventions, we can rewrite the holographic free energy in \cite{Gabella:2012rc} in our convention \eqref{eq:ansatzMetric} of $G_2$ saddles:
	\begin{equation} \label{eq:FreeEnergyHoloG2}
			F_{G_2} = \frac{2^2 5^{\frac{1}{2}} \pi^{\frac{3}{2}}}{3^{\frac{5}{2}}} \sqrt{ \frac{g_0^3}{\int e^{2f_0 + f_1 + 6f_2} d\theta } } =  \frac{2^{\frac{5}{4}} 5^{\frac{3}{4}} \pi^{\frac{3}{2}}}{3^{\frac{5}{2}}} \sqrt{ \frac{g_0^3}{\int e^{3f_0 + 6f_2} d\theta } },
	\end{equation}
	where we have plugged in the convention $e^{f_1} = \frac{2^{3/2}}{5^{1/2}} e^{f_0}$ in the second equality. For the dWNW solution given in \eqref{eq:dWNWSolution}, the integral gives
	\begin{equation}
		I_1 \equiv \int_0^\pi  e^{3f_0 + 6f_2}\Big|_{\rm dWNW} d\theta = \frac{2^{\frac{13}{2}} 3^{\frac{3}{2}} }{ 5^{\frac{7}{2}} } \pi g_0^3 \approx 5.286 g_0^3 ,
	\end{equation}
	which gives the correct value of $F_{G_2}$ as summarized in Table \ref{tbl:HoloFreeEnergy} \cite{Warner:1983vz, deWit:1983vq}. It indeed lies between the free energy of ABJM theory and the CPW theory.
	For the new isolated solution found in section \ref{subsec:IntegralCurvesBranch2}, we can numerically evaluate:
	\begin{equation} \label{eq:NumericalFG2}
		\frac{F_{G_2'} }{N^{3/2}} \approx 1.45669,
	\end{equation}
	which also lies between $F_{\rm SO(8)} \approx 1.4810$ and $F_{\rm GMPS} \approx 1.4331$! Since the free energy window is much more constraining than the $p = 2$ case, this is a very strong support for our proposal of the RG flow with the new $G_2$ IR fixed point. We collect the values of holographic free energies in Table \ref{tbl:HoloFreeEnergy}.
	
			\renewcommand{\arraystretch}{2}
	\begin{table}[h]
		\centering \label{tbl:HoloFreeEnergy}
		\caption{\rm Collection of the large $N$ leading order free energies of the known fixed points. For a complete list of the scalar potentials of different fixed points for both $4d$ gauged supergravity with SO(8) or ISO(7) gaugings, see table 1 of \cite{Cesaro:2021haf}.}
		\begin{tabular}{c| ccc}
			\hline 
			& $\frac{F}{N^{3/2}}$ & global symmetry & supercharges  \\ \hline 
			AdS$_4 \times S^7$ &$\frac{\sqrt{2}\pi}{3}\approx 1.4810$ & SO(8) & 32  \\
			dWNW \cite{deWit:1984nz} & $ \frac{5^{5/2}\pi}{2^2 3^{13/4}} \approx 1.2356 $ & $G_2$ & 4 \\
			 CPW \cite{Corrado:2001nv} & $ \left(\frac{2}{3}\right)^{5/2} \pi \approx 1.1400 $ & SU(3)$\times$ U(1)  & 8 \\ \hline 
			 New (\cite{Duboeuf:2024tbd} and here)  &  $ \approx 1.4567$  & $G_2$ & 4 \\ 
			GMPS \cite{Gabella:2012rc,  Halmagyi:2012ic} & $ \frac{2^6 \pi }{ 3^{9/2} } \approx 1.4331 $ & SU(3)$\times$ U(1)  & 8 \\ \hline 
		\end{tabular}
	\end{table}
	\renewcommand{\arraystretch}{1}
	
	From the table, the expressions of the closed-form planar free energies show some similar patterns. They all involve a factor of $\pi$, with fractional powers of 2, 3, and 5. It is natural to conjecture that the closed-form expression of the planar free energy of the new $G_2$ saddle, if exists, should follow the same pattern. After some numerology, we find that the following expression fits the best with the numerical value \eqref{eq:NumericalFG2}:
	\begin{equation}
		\frac{\pi}{3^{1/3} 5^{1/4}} \approx 1.456689.
	\end{equation}
	
\section{Outlook}
\label{sec:ccl}

By studying the BPS condtions of 11 dimensional supergravity, we identify the most general set of $G_2$-invariant supersymmetric AdS$_4$ backgrounds including a new $G_2$ vacuum. Our study of the holographic free energy indicates that it is the holographic dual of the IR fixed point coming from the cubic deformation to the ABJM theory. Our analysis points to several interesting open questions and directions for future work which we now briefly discuss.

\begin{itemize}

\item With the holographic free energy of the new $G_2$ solution, we give a non-trivial prediction to the study on the field theory side, which is hard because of the low amount of supersymmetries that limits the application of supersymmetric localisation. It would be interesting to understand the nature of the field theory dual of the new bulk saddle. For example, how many independent cubic deformation can we construct? What is the natural of the RG flow from the $G_2$ to the SU(3)$\times$U(1) fixed point? Is this the full web of RG flows from cubic superpotential deformations? We expect a picture similar to that discussed in \cite{Bobev:2009ms} and would like to come back to these questions in the future.

\item It is tantalizing to study the BPS equations analytically and get a closed form expression for the new $G_2$-invariant saddle. This won't be made possible by a brute-force resummation of the series expansion, as the dWNW solution \cite{deWit:1984nz} is found by making a clever ansatz based on the structure of the saddle. We need a clever choice of $f_1$ (perhaps similar to the ones in \eqref{eq:f1Convention}) and also an educated ansatz for the other functions. 

\item The logarithmic correction to the holographic free energy is an interesting quantity to study, which is of order $\log N$ in comparison to the leading term \eqref{eq:Fleading} of order $N^{3/2}$. It can be evaluated by summing over the one-loop contributions of all the Kaluza-Klein modes of the low-energy effective theory. \cite{Bobev:2023dwx} Recently, we make an interesting observation that the logarithmic corrections of different AdS$_4$ Kaluza-Klein supergravity theories are identical if they are connected by a holographic RG flow. \cite{Zhang:2024hrc} It would be meaningful to check this explicitly for the new $G_2$ background and compare it with the AdS$_4 \times S^7$ value \cite{Bhattacharyya:2012ye}, as it would give further support for the web of RG flows picture we propose in this work. A powerful tool along this line is the exceptional field theory method \cite{Malek:2019eaz, Malek:2020yue, Cesaro:2020soq} which calculates the mass spectra of the Kaluza-Klein modes with spin $0 \le s \le 2$. The lack of an analytical form would make this task challenging, but the partial knowledge on some lower Kaluza-Klein levels would already be helpful for the comparison. 

\item We identify an infinite family of $G_2$-invariant saddles in section \ref{subsec:Branch1} whose internal space has the topology of $S^1 \times S^6$. It would be desirable to explore further field theory criteria that select how many of them are holographic, and the nature of their field theory duals.

\item In this work, we study the behavior of the Killing spinors on the $G_2$ invariant saddles as an intermediate step to get the BPS conditions. It is interesting to study this further. For example, whether the spinor can be written in the compact factorised form \eqref{eq:KSCompactForm}, and whether the factor on $S^6$ is related to \eqref{eq:KSonS6}.

\item The spinor projectors encode useful information on the supersymmetry of the system, it would be interesting to find them explicitly. Since our saddle preserves 4 out of 32 real supercharges, we need three projectors in total. The first two can be easily obtained from \eqref{eq:SpinorProjectorsRaw} by restoring the dependence of the Killing spinor on the internal manifold \eqref{eq:S6dependence}, but the third projector along the worldvolume of the M2-branes remains obsecure. It may be of the form $\mathbf{1} - \Gamma^{\textcolor{red}{234}}$ rotated by coordinate-dependent combinations of Gamma matrices as the one in \cite{Pilch:2015vha}.

\end{itemize}

\bigskip

\noindent\textbf{Acknowledgments}

\medskip

\noindent I would like to thank Robert Walker for the collaboration at an early stage of the project. I am grateful to Pieter-Jan de Smet, Vasko Dimitrov, Gabriel Larios, Colin Sterckx, Xinan Zhou for useful discussions, and especially Nikolay Bobev for much help during the project and helpful comments on the draft. XZ is supported by the Fundamental Research Funds for the Central Universities NSFC NO. 12175237, and by funds from the University of Chinese Academy of Sciences. XZ is also supported in part by FWO projects G003523N, G094523N, and G0E2723N, as well as by the Odysseus grant G0F9516N from the FWO. 

\appendix
\addtocontents{toc}{\protect\setcounter{tocdepth}{1}}

\section{Relevant deformation invariant under global symmetry}
\label{App:G2singlet}

   As a primary evidence for the existence of IR fixed points with certain isometries, we show that the relevant deformation operators over ABJM contain singlet sectors under the isometry group to be preserved along the RG flow. We first show that the quadratic and cubic relevant deformations both have a singlet under $G_2$ group, which will induce an RG flow preserving the $G_2$ isometry. We illustrate the same for SU(3)$\times$U(1), which is consistent with the existence of SU(3)$\times$U(1) invariant fixed point.

	   \subsection{$G_2$ isometry}

	The goal is to argue that there is a single quadratic superpotential deformation of ABJM that preserves $\mathcal{N}=1$ supersymmetry and G$_2$ global symmetry. This deformation of the superpotential has the schematic form (in $\mathcal{N}=1$ superspace)
	\begin{equation}
		\Delta W = \text{Tr} XX
	\end{equation}
	with some appropriate G$_2$ invariant contraction of the 8 chiral superfields $X$. To construct an argument along these lines we will use Table 1 of \cite{DHoker:2000pvz}. The quadratic deformations decomposed in components correspond to the following representations of $\SO(8)$ with the corresponding branching under G$_2$ (see for instance \cite{Yamatsu:2015npn})
	\begin{equation}
		\begin{aligned}
			\bf{35_v} &\to  \bf{27} \oplus \bf{7} \oplus {\Red \bf{1}}\,,\\
			\bf{56_s} &\to \bf{27} \oplus \bf{14}\oplus \bf{7}\oplus \bf{7} \oplus {\Red \bf{1}}\,,\\
			\bf{35_c} &\to \bf{27} \oplus \bf{7} \oplus {\Red \bf{1}}\,.
		\end{aligned}
	\end{equation}
	The first line correspond to scalar operators of $\Delta=1$ composed out of 2 elementary scalar fields, the second line is a spin-1/2 operator with $\Delta=3/2$ composed out of an elementary scalar and a spinor, and the third line is a fermionic bilinear with $\Delta=2$. Clearly there is precisely one singlet under G$_2$. This is the deformation that ultimately leads to the AdS$_4$ G$_2$ invariant vacuum of 4d $\mathcal{N}=8$ supergravity \cite{Warner:1983vz}. 
	
	We now proceed in a similar fashion to study a cubic deformation of the schematic form 
	\begin{equation}
		\Delta W = \text{Tr} XXX \,.
	\end{equation}
	The cubic deformations correspond to the following representations of $\SO(8)$ with the corresponding branching under G$_2$
	\begin{equation}
		\begin{aligned}
			\bf{112'} &\to  \bf{77} \oplus\bf{27} \oplus \bf{7} \oplus {\Red \bf{1}}\,,\\
			\bf{224_{vc}} &\to \bf{77} \oplus \bf{64}\oplus \bf{27}\oplus \bf{27} \oplus \bf{14}\oplus \bf{7}\oplus \bf{7} \oplus {\color{red} \bf{1}}\,,\\
			\bf{224_{cv}} &\to \bf{77} \oplus \bf{64}\oplus \bf{27}\oplus \bf{27} \oplus \bf{14}\oplus \bf{7}\oplus \bf{7} \oplus {\red \bf{1}}\,.
		\end{aligned}
	\end{equation}
	The first line correspond to scalar operators $\phi\phi\phi$ of $\Delta=3/2$ composed out of 3 elementary scalar fields, the second line is a spin-1/2 operator $\lambda\phi\phi$ with $\Delta=2$ composed out of 2 elementary scalars and a spinor, and the third line is a $\Delta=5/2$ operator of the form $\lambda\lambda\phi$. Clearly there is precisely one singlet under G$_2$. This suggests that there may be a new AdS$_4$ $\mathcal{N}=1$ vacuum of 11d supergravity with G$_2$ invariance. Notice that since the cubic deformations above lie outside of the 4d $\mathcal{N}=8$ supergravity truncation this should be a genuinely new 11d solution.
	
	   \subsection{SU(3)$\times$U(1) isometry}
	
	There are a few different ways to decompose SO(8) to SU(3)$\times$U(1), but the specific choice does not affect the final conclusion. Suppose the U(1) charges for the decomposed $[SU(3) \times U(1)_1] \times U(1)_2$ are $r_1$ and $r_2$ respectively, the R-charges corresponding to the quadratic and cubic deformations are given by: $$ \small r = \left( \frac{1}{6} - \frac{2}{3p}\right)r_1 + \frac{1}{2} r_2 = \left\{ \begin{aligned}
		&- \frac{1}{6}r_1 + \frac{1}{2}r_2,\quad p = 2;\\
		&   - \frac{1}{18}r_1 + \frac{1}{2}r_2,\quad p = 3.\\
	\end{aligned} \right. $$
	
	We start with quadratic operators, including bosonic ones $X^2, \lambda \lambda$, and fermionic one $\lambda X$. Explicitly using \cite{DHoker:2000pvz}, we have:
	\begin{equation}
		\begin{aligned}
			&X^2:\quad (2000)=\mathbf{35_v}\quad \rightarrow \quad \mathbf{1}_{-2} + \mathbf{1}_{ 0 } + \mathbf{1}_{2} + \cdots, \\
			&\lambda \lambda:\quad (0020)=\mathbf{35_{c} } \quad \rightarrow \quad  \mathbf{1}_{0 } + \mathbf{1}_{0 } + \textcolor{red}{\mathbf{1}_{0 }} + \cdots, \\
			&\lambda X:\quad (1010)=\mathbf{56_{s}} \quad \rightarrow \quad  \mathbf{1}_{ -1 } + \mathbf{1}_{ -1 }  + \mathbf{1}_{ 1 } + \mathbf{1}_{ 1 } +  \cdots, \\
		\end{aligned} 
	\end{equation}
	where $\cdots$ denotes multiplets that are not singlets under the global symmetry group. $X^2$, $\lambda \lambda$ and $\lambda X$ correspond to the scalars, pseudoscalars, and fermions in massless $\cN = 8$ multiplet. In the IR theory, they need to be organized into OSp(2$|$4)$\times$SU(3) multiplets of KK level $n = 0$, if not eaten by Higgs mechanism. The SU(3) singlets form the massive vector multiplet \cite{Klebanov:2008vq} including one singlet $\mathbf{1}_0$ coming from the $\cN = 8$ vectors transferred in $\mathbf{28}$ under SO(8), while the red scalar $\mathbf{1}_0$ above from the pseudoscalar is eaten by it to give it mass.
	
	For the cubic deformation, we have the scalar operators $X^3, \lambda \lambda X$ and the fermionic one $\lambda X^2$, their representations contain the following SU(3) singlets:
	\begin{equation}
		\begin{aligned}
			&X^3:\quad (3000)=\mathbf{112_v} \quad \rightarrow \quad \mathbf{1}_{-2} + \mathbf{1}_{-\frac{2}{3} } + \mathbf{1}_{ \frac{2}{3} } + \mathbf{1}_{2} + \cdots, \\
			&\lambda \lambda X:\quad (1020)=\mathbf{224_{cv}} \quad \rightarrow \quad  \mathbf{1}_{ -\frac{4}{3} } + \mathbf{1}_{ -\frac{2}{3} }  + \mathbf{1}_{ 0 } + \mathbf{1}_{ \frac{2}{3} }   + \mathbf{1}_{ \frac{4}{3} }  + \cdots, \\
			&\lambda X^2:\quad (2010)=\mathbf{224_{vc} }\quad \rightarrow \quad  \mathbf{1}_{ -\frac{5}{3} } + \mathbf{1}_{ -1 }  + \mathbf{1}_{ - \frac{1}{3} } + \mathbf{1}_{ \frac{1}{3} } + \mathbf{1}_{ 1 }   + \mathbf{1}_{ \frac{5}{3} } +  \cdots. \\
		\end{aligned}
	\end{equation}
	A part of the SU(3) singlets form the long vector multiplet of KK level $n = 1$, and a part of them invade $n = 0$ and also forms a long vector multiplet. \cite{Cesaro:2020piw}

\section{The $G_2$ invariant tensors on $S^6$}
\label{App:G2tensors}

	Here we briefly review the construction of the $G_2$ invariant tensors on $S^6$ that are used in the main text. Our presentation closely follows section 2 of \cite{Gunaydin:1983mi}. $J_{mn}$ is the almost complex form on $S^6$, whose standard construction is achieved by embedding $S^6$ as the unit sphere in imaginary octonion ${\rm Im}\, \mathbb{O}$ isomorphic to $\mathbb{R}^7$: \cite{konstantis2018almost, bryant2014s, ballmann2006lectures}
	\begin{equation}
	J_p(v) = J(p, v)\equiv p\times v,\quad v\in T_pS^6, \quad p \in {\rm Im}\, \mathbb{O},
	\end{equation}
	where the product is defined as $u\times v\equiv {\rm Im}(uv) = \frac{1}{2}(uv - vu)$. Consider the imaginary octonion $ {\rm Im}\, \mathbb{O} \simeq \mathbb{R}^7$ has a basis $\{e_1, e_2,\cdots ,e_7\}$, the almost complex structure $J_p$ on $S^6$ for a given $p = p^\mu e_\mu \in S^6$ is given by: \footnote{$\mu,\nu,\rho, \cdots $ are flat indices on $\mathbb{R}^7$ where $S^6$ is embedded, ranging from 1 to 7.}
	\begin{equation}
	J_{\ \nu}^{\rho}v^\nu e_\rho \equiv J_p v = \frac{1}{2}(pv-vp) =p^\mu v^\nu \tilde{\eta}_{\mu\nu}^{\ \ \rho} e_\rho,
\end{equation}
	where $\tilde{\eta}_{\mu\nu}^{\ \ \rho}$ is defined in the inner product of octonions:\footnote{The convention of $\eta_{[\mu\nu}^{\quad \rho]}$ varies in the literature, what we take is consistent with \cite{Gunaydin:1983mi}.} \footnote{We are sloppy on the distinction of upper and lower indices, but $\tilde{\eta}$ is completely anti-symmetric in its three indices.} 
	\begin{equation}
	e_\mu e_\nu = -\delta_{\mu\nu}e_0 + \tilde{\eta}_{[\mu\nu}^{\quad \rho]}e_\rho, \quad \tilde{\eta}_{[\mu\nu}^{\quad \rho]}=1\ {\rm when}\ \mu\nu\rho = 123, 471, 572, 673, 354, 246, 165.
	\label{OctonicStructureConstant}
\end{equation}
	 From the above equation we get:
	\begin{equation}
	J_{\rho\nu}(p) = p^\mu \tilde{\eta}_{\mu\nu\rho}.
\end{equation}
	Note that the indices $\mu,\nu,\rho$ are on $\mathbb{R}^7$, and $m,n,p$ are on $S^6$, we assume the coordinate transformation between they two to be:
	\begin{equation}
	dx^\mu = \frac{\partial x^\mu}{\partial x^m} dx^m \equiv \Lambda^\mu_{\ m} dx^m,
\end{equation}
	where the form of transformation $\Lambda_{\ m}^\mu$ is easily obtained once a coordinate on $S^6$ is set, and we have
	\begin{equation}
	J_{mn}(p) = J_{\rho\nu}(p)\Lambda_{\ m}^\rho \Lambda_{\ n}^\nu = p^\mu \tilde{\eta}_{\mu\nu\rho}\Lambda_{\ m}^\rho \Lambda_{\ n}^\nu,
\end{equation}
	expressed in the coordinate on $S^6$. The above definition is equivalent to that in \cite{Gunaydin:1983mi}, which is:
	\begin{equation}
	J_\mu^{\ \nu}e_\nu = e_\mu \times p.
\end{equation}
	In practice, we took the most na\"ive coordinate on $S^6$:
	\begin{equation}
	\begin{aligned}
		& x^1 = \cos\psi^1, &\\
		& x^2 = \sin\psi^1 \cos\psi^2,  &\\
		& x^3 = \sin\psi^1 \sin\psi^2 \cos\psi^3, &\\
		&......&\\
		& x^{6} = \sin\psi^1 \sin\psi^2 \sin\psi^3\sin\psi^4 \sin\psi^5 \cos\psi^6,  &\\
		& x^{7} = \sin\psi^1 \sin\psi^2 \sin\psi^3\sin\psi^4 \sin\psi^5 \sin\psi^6. &\\
	\end{aligned}
\end{equation}
	
	The tensor $T_{\mu\nu\rho}$ is obtained by definition:
	\begin{equation}
	T_{\mu\nu\rho} = (e_\mu \times e_\nu, e_\rho) = \tilde{\eta}_{\mu\nu\rho},\quad \Rightarrow \quad T_{mnp} = \Lambda^{\mu}_{\ m} \Lambda^{\nu}_{\ n} \Lambda^{\rho}_{\ p}\tilde{\eta}_{\mu\nu\rho},
\end{equation}
	where the definition of inner product $(\cdot, \cdot)$ is the same as that for vectors in $\mathbb{R}^7$:
	\begin{equation}
	(e_\mu, e_\nu) \equiv e_\mu^\alpha e_{\nu\alpha}.
\end{equation}
	The dual of the torsion tensor is defined as:
	\begin{equation}
		S_{mnp} \equiv \frac{1}{3!} \epsilon_{mnpqrs} T^{qrs}.
	\end{equation}
	For a more modern formalism of the almost complex form in the context of supergravity, see for example \cite{Larios:2019kbw}.
	
	\section{A detailed study of the dynamical system}
	\label{App:DynamicalSystem}
	
	We will present a detailed analysis of the BPS equations \eqref{eq:BPSEquations} as two dynamical systems. In section \ref{subsec:f2saddle}, we will present the families of integral curves that start from the extremal point of $f_2(\theta)$, following the preliminary analysis of section \ref{StationaryLocus}. In section \ref{sec:ConnectingFamilies}, we will demonstrate that none of the combinations of integral curves, except for the one presented in the maintext, give a smooth $G_2$ saddle we are after.
	
	\subsection{Series expansion at the saddle of $f_2$ and numerics: two Branches}
	\label{subsec:f2saddle}
	
	In section \ref{sec:G2Saddles}, we only present the families of integral curves relevant for the new $G_2$ saddle. Here, we will present all the others that start from the extremal point of $f_2(\theta)$. As discussed in section  \ref{StationaryLocus}, the set of extremal points of $f_2$ includes Branch 1, Branch 2, and a two-dimensional set. There are three families of integral curves starting from Branch 1, also three starting from Branch 2, including the one discussed in section \ref{subsec:IntegralCurvesBranch2}, and only one starting from the two-dimensional set. We discuss them seperately in what follows.
	
	\subsubsection{Branch 1: $\hat{f}_{2,0} = 6 \hat{f}_{0,0}^4$}
	\label{subsec:Branch1}
	
	We start with Branch 1. Plugging the leading-order relation into the BPS equations and solving them order-by-order, we get three different families of solutions, each has one free parameter $\hat{f}_{0,0} = e^{f_0}(\theta_*)$. One of them includes the dWNW solution:
	\begin{equation}   \label{eq:ExpansionfromMiddle}
		\begin{aligned}
			e^{f_0} &= \hat{f}_{0,0} + \frac{4\left( 9 \hat{f}_{0,0}^6 - 1 \right) (\theta - \theta_*)^2 }{9 \hat{f}_{0,0}^5} + \frac{8\left( 9 \hat{f}_{0,0}^6 - 1 \right) \left(  195 \hat{f}_{0,0}^6-29 \right) (\theta - \theta_*)^4 }{ 405\hat{f}_{0,0}^{11} } + O(\theta - \theta_*)^6, \\
			e^{f_2} &= 6 \hat{f}_{0,0}^4 -  \frac{2 \left( 180 \hat{f}_{0,0}^6 - 19 \right) (\theta - \theta_*)^2 }{15 \hat{f}_{0,0}^2} - \frac{\left( 75600  \hat{f}_{0,0}^{12}-20760  \hat{f}_{0,0}^6+1373\right)(\theta - \theta_*)^4 }{675 \hat{f}_{0,0}^8} + O(\theta - \theta_*)^6,\\
			\tilde g_3 &= 6^6  \hat{f}_{0,0}^{24} \left( 9 \hat{f}_{0,0}^6 - 1 \right) \left[ 1 - \frac{16 (\theta - \theta_*)^2 }{15 \hat{f}_{0,0}^6 }  + \frac{8 \left(340 \hat{f}_{0,0}^6-27\right)(\theta - \theta_*)^4 }{225 \hat{f}_{0,0}^{12}} + O(\theta - \theta_*)^6 \right]. 
		\end{aligned}
	\end{equation}
	The condition $\tilde g_3 \ge 0$ dictates $\hat{f}_{0,0} \ge \frac{1}{3^{1/3}} \approx 0.693$. The dWNW solution corresponds to $\hat{f}_{0,0} = \left( \frac{2}{15} \right)^{1/6} \approx 0.715$. With the initial conditions around $\theta = \theta_*$, we try to numerically integrate the BPS equations, but the result is not very well. Since it includes the dWNW solution, this family belongs to \DS1.
	
\begin{figure}[h]
		\centering \caption{\rm {\it Left panel}: the second family of solutions extrapolated by Pad\'e. {\it Right panel:} the third family obtained by numerics. They are lying on the surfaces where $f_0'$ and $f_2'$ begin to be imaginary. The blue curve is the dWNW solution. }
		\includegraphics[width=7cm]{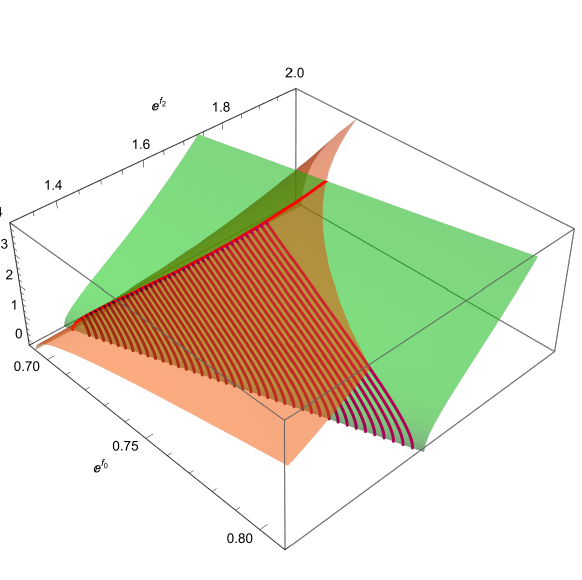}
		\includegraphics[width=7cm]{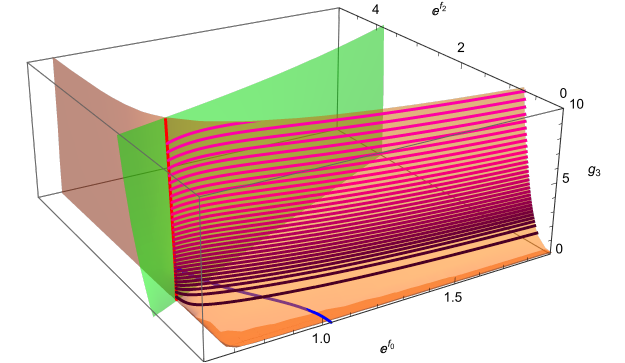}
		\label{fig:MiddleFamilies}
	\end{figure}

\begin{figure} [!h]
		\centering 
		\caption{\rm We show the periodic solution with initial value $\hat{f}_{0,0} = 0.7$, with the three plots $e^{f_0}$, $e^{f_2}$, and $g_3$ as functions of $\theta$. In this example, $f_0$ and $f_2$ are very close to constants. } 
		\includegraphics[width=4.5cm]{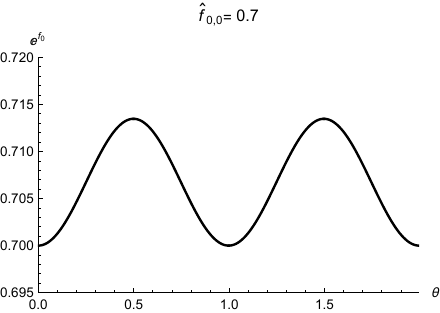} \hspace{0.3cm}
		\includegraphics[width=4.5cm]{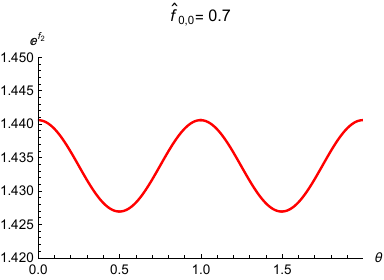} \hspace{0.3cm}
		\includegraphics[width=4.5cm]{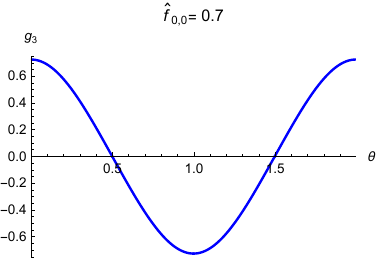}
		\label{fig:Periodic solutions}
	\end{figure}

	The second family admits the following expansion: \footnote{Notice that we initially do the $(\theta - \theta_*)$ expansion instead of the $\rho \equiv (\theta - \theta_*)^2$ expansion. It is interesting that only the even terms of $(\theta - \theta_*)$ appear. This may have to do with the invariance of the BPS equations in terms of integration direction we mention before. }
	\begin{equation}  \label{eq:ExpansionFamilyGreen}
		\begin{aligned}
			e^{f_0} &= \hat{f}_{0,0} + \frac{2\left( 9 \hat{f}_{0,0}^6 - 1 \right)\rho }{45 \hat{f}_{0,0}^{11} } - \frac{4\left( 9 \hat{f}_{0,0}^6 - 1 \right)\rho^2 }{ 2025 \hat{f}_{0,0}^{23} } + O(\rho)^3,\quad \rho \equiv (\theta - \theta_*)^2,\\
			e^{f_2} &= 6 \hat{f}_{0,0}^4 -  \frac{2 \left( 9 \hat{f}_{0,0}^6 - 1 \right)\rho }{15 \hat{f}_{0,0}^8 } + \frac{\left( 243  \hat{f}_{0,0}^{12}- 18  \hat{f}_{0,0}^6 -1 \right) \rho^2 }{675 \hat{f}_{0,0}^{20} } + O(\rho)^3, \\
			\tilde g_3 &= 6^6  \hat{f}_{0,0}^{24} \left( 9 \hat{f}_{0,0}^6 - 1 \right) \left[ 1 - \frac{2 \left( 18 \hat{f}_{0,0}^6 - 1 \right) \rho}{15 \hat{f}_{0,0}^{12} }  + \frac{4 \left(567 \hat{f}_{0,0}^{12} - 63 \hat{f}_{0,0}^{6} + 1 \right) \rho ^2}{675 \hat{f}_{0,0}^{24}} + O(\rho)^3 \right],
		\end{aligned}
	\end{equation}
	where we have introduced $\rho \equiv (\theta - \theta_*)^2$ to simplify the notation. Numerics is not very stable around the saddle point, so we evaluate the expansion up to $O(\theta - \theta_*)^{30}$ and use Pad\'e approximation to extrapolate the behavior. As shown in the left panel of Fig. \ref{fig:MiddleFamilies}, this family of integral curves completely lies on the green surface whose expression is given by:
	\begin{equation} \label{eq:GreenSurface}
		g_3 = e^{3f_2} \sqrt{ \frac{1}{4}e^{-2f_0 + 2f_2} -1 },
	\end{equation}
	on which the discriminant of the quadratic algebraic equations for $f_0', f_2'$ vanishes. 
	With this observation, we constraint the differential equations on the surface and reduce the BPS equations. Interestingly, one of them is simple:
	\begin{equation}
		f_0'(\theta) + 2 f_2'(\theta) = 0 \quad \Rightarrow \quad e^{f_2} = C e^{-\frac{1}{2}f_0}.
	\end{equation}
	Another property of this family of integral curves is that they run periodically between Branch 1 and Branch 2. With the simplified BPS equations, we can do precise numerics and identify infinitely many new solutions periodic in $\theta$, one of them is shown in Fig. \ref{fig:Periodic solutions}. Unfortunately, the internal space has the topology of $S^1 \times S^6$, not of $S^7$ that we are after. Because of the different topology, we don't expect their dual field theories, if any, to be connected to ABJM by an RG flow. It would be interesting to understand their holographic dual.
	
	Now we move to the third family, for which the expansion around $\theta = \theta_*$ is:
	\begin{equation} \small  \label{eq:ExpansionFamilyOrange}
		\begin{aligned}
			e^{f_0} &= \hat{f}_{0,0} + \frac{2\left( 9 \hat{f}_{0,0}^6 - 1 \right)\left( 10 \hat{f}_{0,0}^6 - 1 \right)\rho }{45 \hat{f}_{0,0}^{11} } - \frac{2\left( 9 \hat{f}_{0,0}^6 - 1 \right) \left( 100 \hat{f}_{0,0}^{12}  - 1 \right) \left( 39 \hat{f}_{0,0}^6 - 4 \right) \rho^2 }{ 2025 \hat{f}_{0,0}^{23} } + O(\rho)^3, \\
			e^{f_2} &= 6 \hat{f}_{0,0}^4 -  \frac{2 \left( 10 \hat{f}_{0,0}^6 - 1 \right) \left( 18 \hat{f}_{0,0}^6 - 1 \right) \rho }{15 \hat{f}_{0,0}^8 } - \frac{\left( 10 \hat{f}_{0,0}^6 - 1 \right) \left( 7560 \hat{f}_{0,0}^{18} - 348  \hat{f}_{0,0}^{12}- 154  \hat{f}_{0,0}^6 + 11 \right) \rho^2 }{675 \hat{f}_{0,0}^{20} } + O(\rho)^3, \\
			\tilde g_3 &= 6^6  \hat{f}_{0,0}^{24} \left( 9 \hat{f}_{0,0}^6 - 1 \right) \left[ 1 + \frac{2 \left( 10 \hat{f}_{0,0}^6 - 1 \right) \rho}{15 \hat{f}_{0,0}^{12} }  - \frac{4 \left( 10 \hat{f}_{0,0}^6 - 1 \right) \left(120 \hat{f}_{0,0}^{12} - 51 \hat{f}_{0,0}^{6} + 4 \right) \rho ^2}{675 \hat{f}_{0,0}^{24}} + O(\rho)^3 \right].
		\end{aligned}
	\end{equation}
	From the numerical Pad\'e approximation it is easy to see that these integral curves all lie on the degenrate orange surface, whose expression is
	\begin{equation} \label{eq:OrangeSurface}
		g_3 = e^{3f_2} \sqrt{9e^{6f_0} - 1} .
	\end{equation}
	So again, we constrain the differential equations on the surface and solve it numerically for $e^{f_0}, e^{f_2}$. The result is shown in the right panel of Fig. \ref{fig:MiddleFamilies}, from which it is apparent that all the integral curves move along the direction with divergent $e^{f_0}$ and vanishing $e^{f_2}$. As shown in the left panel of Fig. \ref{fig:Family3Orange} below, when the initial value $\hat{f}_{0,0}$ approaches the lowest value $\frac{1}{3^{1/3}}$, the solution reduces to AdS$_4 \times S^7$, consistent with what we see in the 3-dimensional plot. For a generic $\hat{f}_{0,0} > \frac{1}{3^{1/3}}$ such as the right panel of Fig. \ref{fig:Family3Orange}, the solution is singular, and the larger $\hat{f}_{0,0}$ is, the more singular it is. In the limit $\hat{f}_{0,0}\rightarrow  \infty$, the functions $e^{f_0},e^{f_2}$ behave like step functions. We draw the conclusion that all integral curves that qualitatively behave like this will not give a regular solution.
	
	\begin{figure}[h]
		\centering \caption{\rm We show the values of $e^{f_0}$ and $e^{f_2}$ along two of the integral lines in the third family of solutions. The left panel corresponds to the limit $\hat{f}_{0,0} \rightarrow \frac{1}{3^{1/3}} \approx 0.693$, where the solution reduces to AdS$_4 \times S^7$. As explained in footnote \ref{footnote:rangeofTheta}, the range of $\theta$ in our plot is $ \sqrt{\frac{5}{2}}  \frac{\pi}{2} \approx 2.48$. The right panel chooses the initial value $\hat{f}_{0,0} = 0.7$. The range of $\theta$ is finite, while the AdS length scale diverges. } 
		\includegraphics[width=7cm]{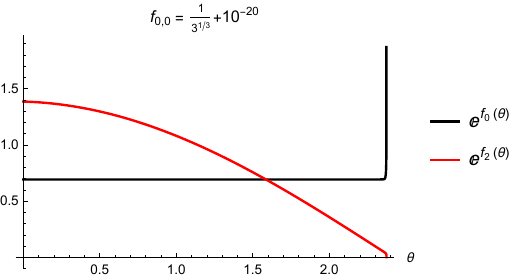} \hspace{0.5cm}
		\includegraphics[width=7cm]{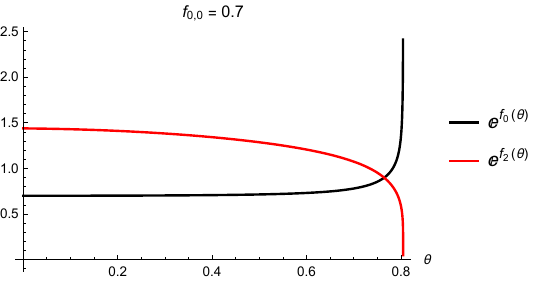}
		\label{fig:Family3Orange}
	\end{figure}
	
	\subsubsection{Branch 2: $\hat{f}_{2,0} = 2 \hat{f}_{0,0}$}
	
	From the power series expansion on Branch 2, we also get three families of solutions, depending on one free parameter $\hat{f}_{0,0} > \frac{1}{3^{1/3}}$. One of them is already presented in section \ref{subsec:IntegralCurvesBranch2}. The second one of them is simple and can be resummed into:
	\begin{equation} \label{eq:ExpansionFamily1Branch2}
		e^{f_2} = 2 \cos \left( \sqrt{\frac{2}{5}} (\theta - \theta_*) \right) e^{f_0}, \quad e^{f_0} \equiv \hat{f}_{0,0}, \quad g_3 = 0,
	\end{equation}
	which is nothing but the AdS$_4 \times S^7$ saddle with a free parameter to be fixed using the trombone symmetry \eqref{eq:trombone}.
	
	The last family of solution has the following expansion:
	\begin{equation} \label{eq:ExpansionFamily2Branch2}
		\begin{aligned}
			e^{f_0} &= \hat{f}_{0,0} - \frac{2\left( 9 \hat{f}_{0,0}^6 - 1 \right)\rho }{15 \hat{f}_{0,0}^{5} } + \frac{4\left( 9 \hat{f}_{0,0}^6 - 1 \right)\left( 18 \hat{f}_{0,0}^6 + 1 \right) \rho^2 }{ 225 \hat{f}_{0,0}^{11} } + O(\rho)^3,\quad \rho \equiv (\theta - \theta_*)^2,\\
			e^{f_2} &= 2 \hat{f}_{0,0} +  \frac{2 \left( 9 \hat{f}_{0,0}^6 - 1 \right)\rho }{15 \hat{f}_{0,0}^5 } - \frac{  \left( 9 \hat{f}_{0,0}^6 - 1 \right) \left( 45 \hat{f}_{0,0}^6 + 7 \right) \rho^2 }{225 \hat{f}_{0,0}^{11} } + O(\rho)^3, \\
			\tilde g_3 &= \frac{128}{5} \left( 9 \hat{f}_{0,0}^6 - 1 \right) \left[ \rho + \frac{2 \left( 9 \hat{f}_{0,0}^6 - 2 \right) \rho^2 }{5 \hat{f}_{0,0}^{6} }  - \frac{12 \left( 39 \hat{f}_{0,0}^{6} -5 \right) \rho^3 }{125 \hat{f}_{0,0}^{12}} + O(\rho)^4 \right],
		\end{aligned}
	\end{equation}
	which can be shown to reproduce the second family of solutions \eqref{eq:ExpansionFamilyGreen} from Branch 1, i.e., this family of solutions connects Branch 1 and 2.
	
	As a quick summary, starting from the two branches, we get three families of integral curves \eqref{eq:ExpansionFamilyGreen} \eqref{eq:ExpansionFamilyOrange} \eqref{eq:ExpansionFamily1Branch2} lying on the degenerate surfaces, one family of curves \eqref{eq:ExpansionfromMiddle} flowing along \DS1 and another family \eqref{eq:ExpansionFamily3Branch2} along \DS2.
	
		\begin{figure}[!h]
		\centering 
		\caption{\rm We draw two 1-dimensional subsets of the two-dimensional families of integral curves, one is to the left of the dWNW solution (the blue curve appearing in the right panel), and the other is to the right. The black curve is swept by the end points of all integral curves of \DS 1 from $\theta = 0$ (the orange curves in Fig. \ref{fig:IntegrationCurves}). } 
		\includegraphics[width=6cm]{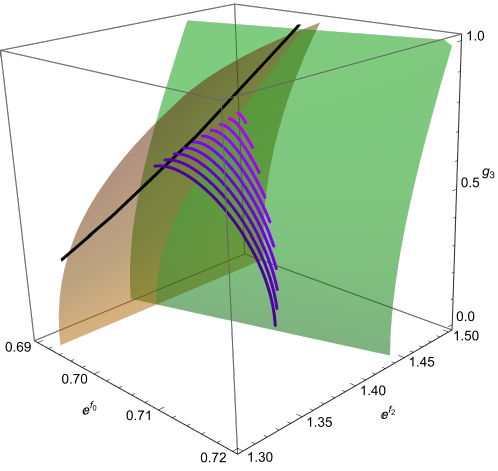} \hspace{0.5cm}
		\includegraphics[width=7cm]{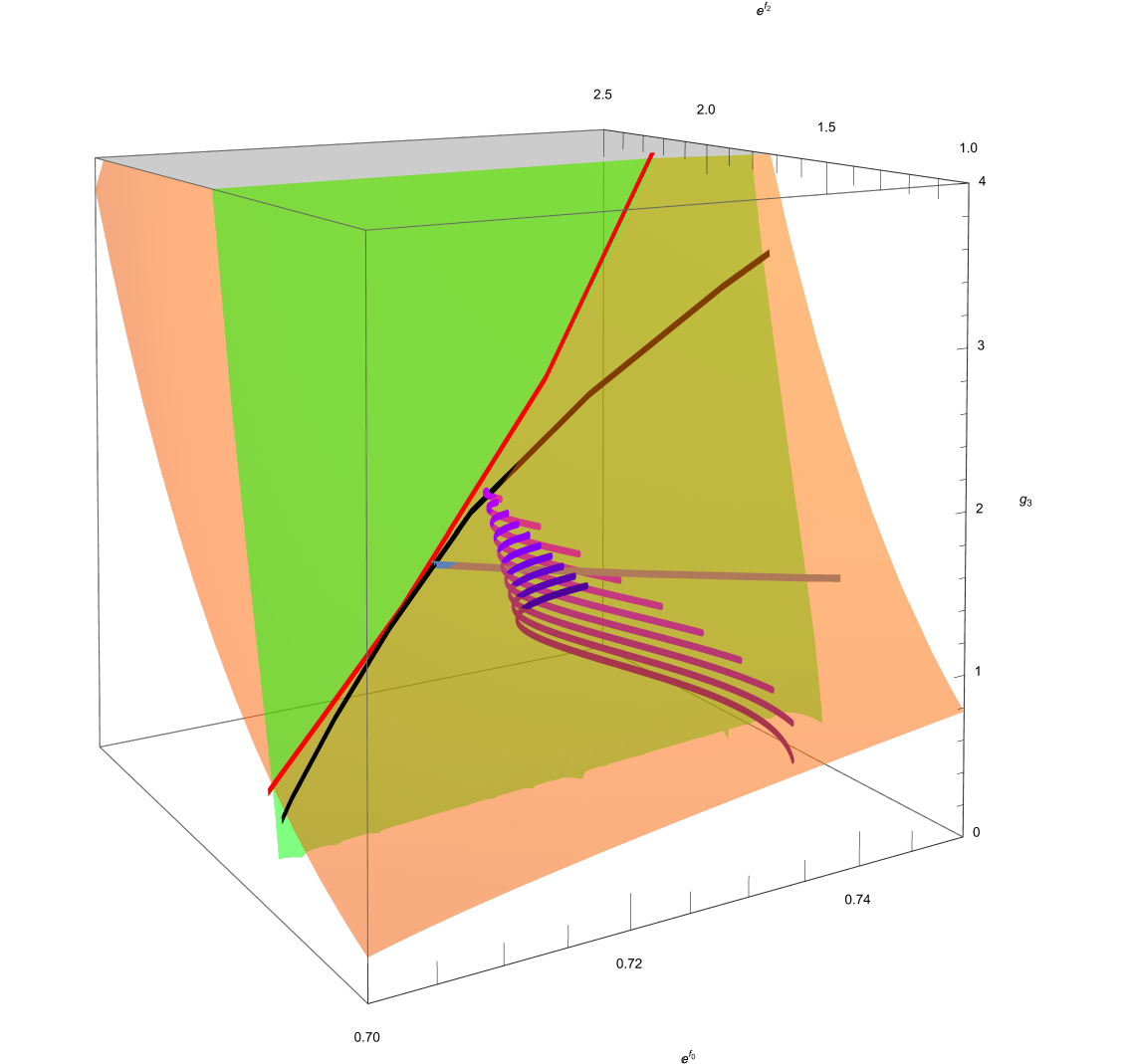}
		\label{fig:IntersectionS1S2}
	\end{figure}
	
	\subsubsection{The two-parameter families}
	
	Now we turn to the two-parameter family of integral curves discussed around \eqref{eq:Solution2}. Since the extremal points of $f_2$ are not stationary points of \DS 2, it is less meaningful to initiate the numerical integration from those points. Instead, we start the numerical integration on points infinitesimally close to the green surface parametrized by two initial values $\alpha \equiv e^{f_0}(\theta_{\rm ini})$ and $\beta \equiv e^{f_2}(\theta_{\rm ini})$, with the initial value of $g_3$ given by \eqref{eq:GreenSurface}. When doing the numerical integral, we also need to take care of the non-analytical behavior around the surface \eqref{eq:NonAnalyticSurface}. In Fig. \ref{fig:IntersectionS1S2}, we present two of the one-dimensional sub-families of integral curves with fixed $\alpha$ and various $\beta$. All the integral curves connect the green and orange surfaces. 

	 For a fixed $\alpha$, the initial value $\beta$ is constrained by $ 2\alpha < \beta < 6\alpha^4$. As $\beta \rightarrow 2\alpha$, the starting point is on Branch 2, and the integral curve is exactly what we discuss in section \ref{subsec:IntegralCurvesBranch2}. In other words, the integral curves are bounded from below by the ones discussed in section \ref{subsec:IntegralCurvesBranch2}. As $\beta \rightarrow 6 \alpha^6$, the length of the integral curve decreases to zero.
	
	 In what follows, we will try to connect integral curves belonging to different families with the goal of constructing a smooth solution.

	\subsection{Connecting different families}
	\label{sec:ConnectingFamilies}
	
	It is easier to view a solution of the BPS conditions as an integral curve in the dymical system. From this perspective, the analysis in section \ref{subsec:theta=0} amounts to telling us that a regular solution has to start on the $e^{f_0}$-axis\footnote{i.e., $e^{f_2} = g_3 = 0$ at the initial point. } with an initial value $e^{f_0}(\theta = 0) = \tilde{f}_{0,0} > \frac{1}{3^{1/3}}$, moves along an integral curve of either \DS 1 or \DS 2, or a smooth combination of integral curves along different Dynamical Systems, and in the end, goes back to the $e^{f_0}$-axis. Take the dWNW solution as an example, we start at $ \tilde{f}_{0,0} = (6/5)^{1/6}$ and flow along \DS 1 until we reach Branch 1. Luckily, this is a stationary point of the dymical system, so we can reverse the system by replacing $(f_0', f_2', g_3') \rightarrow -(f_0', f_2', g_3')$ and still have a smooth solution. The system will evolve backwards and finally reach the $e^{f_0}$-axis. This is illustrated in Fig. \ref{fig:IntegrationCurves}. 
	
	Another necessary condition for a valid family is the existence of $\theta_*$ where $f_2'(\theta_*) = 0$, this means that the integral curve has to reach Branch 1, Branch 2, or the 2-dimensional surface. In what follows, we will systematically discuss all the possibilities and thus search the most general set of $G_2$-invariant saddles in 11 dimensional supergravity.
	
	Starting from the $e^{f_0}$-axis, there is only one family of integral curves \eqref{eq:PowerExpansionFamily4} that flows along \DS1. Purely in \DS1, there is no new solution apart for dWNW and AdS$_4\times S^7$, so we need to transfer to \DS2 somewhere $\theta_t > 0$ at $\left( e^{f_0}(\theta_t), e^{f_2}(\theta_t), g_3(\theta_t) \right)$. We expect the transfer to be smooth, one necessary condition of which is that the quadratic algebraic equations of $f_0'$ and $f_2'$ \eqref{eq:BPSEquations} are degenerate, which only happens on the green and orange surfaces (as well as their intersections). This leads to the conclusion that a smooth transfer to \DS 2 may only happen on the two surfaces.\footnote{In fact, the plane $g_3 = 0$ is also a degenerate surface, whose intersection with the green surface is Branch 2. But except for the fact that the AdS$_4 \times S^7$ saddle lies in this plane, it turns out to be less relevant in our discussion. } In what follows, we will discuss different initial values $\tilde{f}_{0,0}$ and whether they give smooth solutions. 
	
	\subsubsection{ Initial condition $\tilde{f}_{0,0} = \tilde{f}_{0,0}^{\rm dWNW}$ }
	
\begin{figure}
		\centering 
		\caption{\rm An example of the solution that is not smooth everywhere in the range of $\theta$. } 
		\includegraphics[width=7cm]{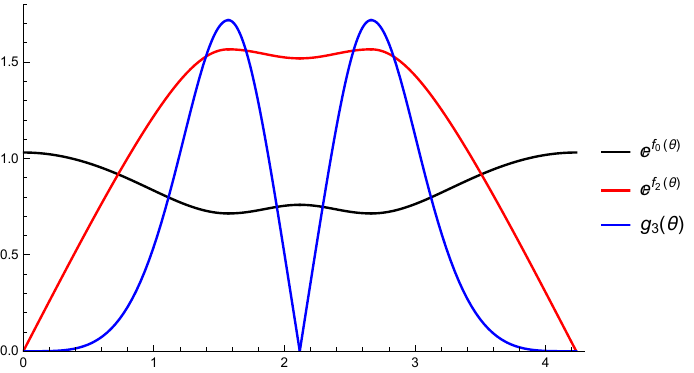}
		\label{fig:GluingSolutionsn=1}
	\end{figure}

	Let's start with the special initial value $\tilde{f}_{0,0} = \tilde{f}_{0,0}^{\rm dWNW}$. When we follow the integral curve of \DS 1 and reach Branch 1, we can choose not to evolve backwards directly but evolve along the integral curves \eqref{eq:ExpansionFamilyGreen} on the green surface, this will give a solution that seems smooth such as the one shown in Fig. \ref{fig:GluingSolutionsn=1}. However, although the first order derivative is smooth, the second order derivative is not, as can be seen from the different subleading terms in the series expansions \eqref{eq:ExpansionfromMiddle} and \eqref{eq:ExpansionFamilyGreen}. 
	
	It may be confusing how could the second order derivative be discontinuous while the first order derivative is continuous, as this seem to violate the equations of motion \eqref{eq:eomComponents}. This is because the first-order derivative is not differentiable at the intersection and thus the BPS equations don't imply the equations of motion by differentiating over $\theta$. As a result, the integral curve doesn't solve the equations of motion at this special value of $\theta$.
	
	\subsubsection{ Initial condition $\tilde{f}_{0,0} > \tilde{f}_{0,0}^{\rm dWNW}$ }
	
\begin{figure}[!h]
		\centering 
		\caption{\rm {\it Left panel:} the connection between the two families of integral curves, with the yellow ones starting from the $e^{f_0}$-axis and the pink ones lying on the green surface. The red curve represents Branch 1 of stationary points. {\it Right panel:} The orange-brown family of curves start from $\theta = 0$ and include the dWNW solution colored in blue. The pink family of curves belong to the third family of solutions \eqref{eq:ExpansionFamilyOrange} lying in the orange surface and stemming from Branch 1. } 
		\includegraphics[width=7cm]{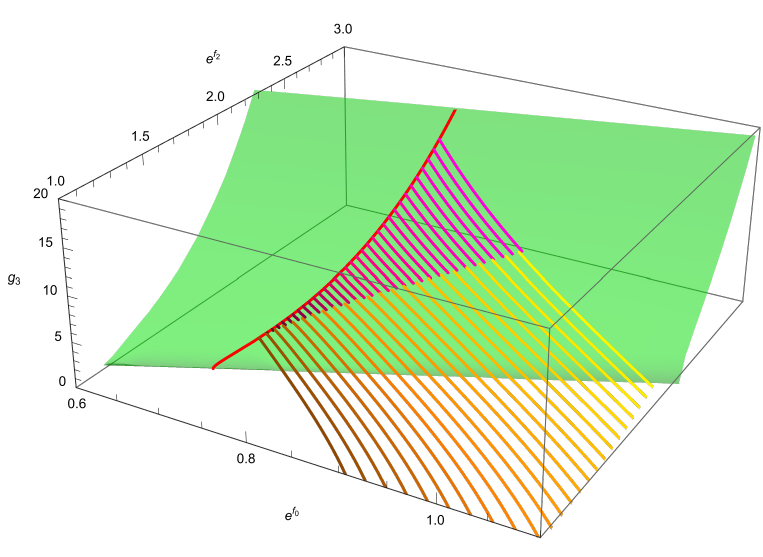} \hspace{0.5cm}
		\includegraphics[width=6cm]{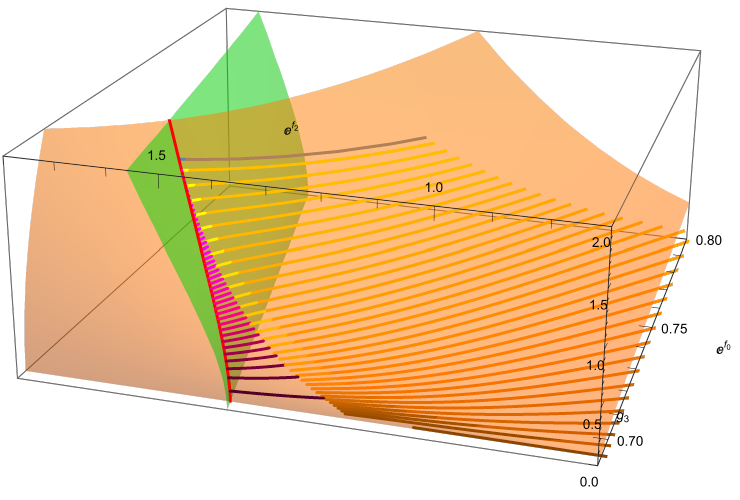}
		\label{fig:GluingSolutions}
	\end{figure}

	As already shown in Fig. \ref{fig:IntegrationCurves} in the maintext, with initial condition $\tilde{f}_{0,0} > \tilde{f}_{0,0}^{\rm dWNW}$, the integral curve reaches the green surface. Furthermore, the integral curve doesn't really stop there, as the dynamical system is well-defined on the green surface as discussed in section \ref{subsec:Branch1}. By combining the two families of integral curves, we get a new family shown in the left panel of Fig. \ref{fig:GluingSolutions}. Since the dynamical system is degenerate on the green surface, the first-order derivatives at the intersection points should be continuous. However, the second-order derivative is discontinuous at the intersection point: the larger $\tilde{f}_{0,0}$ we take, the larger difference we get between the two sides. So we draw the conclusion that the new family of integral curves does not give any smooth solutions.
	
	\subsubsection{ Initial condition $\tilde{f}_{0,0} < \tilde{f}_{0,0}^{\rm dWNW}$ }
	
	In constrast to previous case, with the initial condition $\tilde{f}_{0,0} < \tilde{f}_{0,0}^{\rm dWNW}$, the integral curves approach the orange surface and connect with the family of curves \eqref{eq:ExpansionFamilyOrange}. This generates a new family of curves shown in the right panel of Fig. \ref{fig:GluingSolutions}. Same as previous case, the second-order derivatives on the two sides still don't match. So the conclusion is the same, i.e., there is no smooth solution.
	
	\subsubsection{Two-parameter family in \DS2}
	The last family of integral curves is the one with two parameters $\alpha = e^{f_0}(\theta_{\rm ini})$ and $\beta = e^{f_2}(\theta_{\rm ini})$ with the constraint $\alpha > \frac{1}{3^{1/3}}$ and $2\alpha < \beta < 6\alpha^4$. 
	
	It is hard to search among the full two-parameter family of curves by brute force, so we first reduce irrelevant cases. The integral curves are bounded from below by the ones discussed in section \ref{subsec:IntegralCurvesBranch2}, meaning that all the integral curves lie above the surface swept by the curves shown in Fig. \ref{fig:Branch2Family3}. From there we can see that if we take a large initial value of $\alpha$, the corresponding integral curve will be very close to the orange surface, this asymptotic behavior is analogue to the family of curves on the orange surface shown in the right panel of Fig. \ref{fig:Family3Orange}. As we argued there, this asymptotic behavior only leads to singluar solutions. So we do not need to consider the full non-compact space of initial parameters $(\alpha, \beta)$, but only a finite space bounded by, for example, $\alpha \lesssim 10 \alpha_{\rm dWNW}$. 	
	
	Let us have a look how would a possible solution constructed in this way look like. It belongs to \DS 1 both in the beginning and the end of the interval $I_\theta$, and a sector of \DS 2 is sandwiched between them. Since we have found no continuous connection between \DS 1 and \DS 2 except for the single special case in section \ref{subsec:IntegralCurvesBranch2}, and the integral curves in \DS 2 are not intersecting with each other, we expect a single integral curve to be sandwiched between two curves of \DS 1. It is not hard to see that this does not happen, as shown by the two representative examples in Fig. \ref{fig:IntersectionS1S2}: the integral curves of \DS 2 are U-shaped and connect the green and orange surfaces. None of them intersect the black curve on the two sides at the same time, so none of them give a smooth solution. 
	
	To summarize this section, we have exhausted all possibilities to combine the integral curves between \DS 1 and \DS 2, and the only non-trivial connection is the one presented in section \ref{subsec:IntegralCurvesBranch2}, which gives a brand new $G_2$-invariant solution. Besides, we discover an infinite famiy of solutions periodic in $\theta$, whose internal space topology is $S^1 \times S^6$. Our analysis also excludes the existence of any other supersymmetric $G_2$-invariant saddles in 11d supergravity, including the one denoted by $G_2''$ in \cite{Duboeuf:2024tbd}.
	
	\section{Comparison with \cite{Duboeuf:2024tbd}}
	\label{App:CompareDuboeuf}
	
	Here we compare our results with \cite{Duboeuf:2024tbd} and give evidence why is the solution named $G_2''$ there non-supersymmetric. The map betwen our notations is:
	\begin{equation} \label{eq:Mapping}
		g_0 = \frac{1}{4} F \left( \frac{\ell_4}{\ell_4^{(0)}}\right)^4,\quad e^{f_0} = \frac{1}{2} \Delta^{-\frac{1}{2}} \frac{\ell_4}{\ell_4^{(0)}}, \quad e^{f_1} = e^{\frac{3}{2} \phi} \Delta^{-\frac{1}{2}},\quad e^{f_2} = e^{-\phi/4} \Delta^{1/4} \sin\theta,
	\end{equation}
	where $\ell_4$ is the AdS length scale, and $\ell_4^{(0)}$ is a reference quantity with mass dimension $-1$ which does not change under the trombone symmetry \eqref{eq:trombone}, while the other quantities transform as:
	\begin{equation}
		e^\phi \rightarrow \lambda^3 e^\phi,\quad \Delta \rightarrow \lambda^7 \Delta, \quad A \rightarrow \lambda^3 A,\quad F \rightarrow \lambda^{-15}F,\quad \ell_4 \rightarrow \lambda^{7/2} \ell_4.
	\end{equation}
	To do numerics, we gauge the trombone symmetry by fixing $g_0 = 1$, while in \cite{Duboeuf:2024tbd} they fix $F = \frac{3}{2}$. The map between our conventions cannot be realised because our ignorance to the dimensionless ratio $\ell_4/\ell_4^{(0)}$. Nevertheless, the holographic free energy \eqref{eq:FreeEnergyHoloG2} gives us a hint, for \eqref{eq:Mapping} and $F = 3/2$ it gives
		\begin{equation} \label{eq:holoF}
		\frac{F_{\rm holo}}{N^{3/2}} = \frac{\sqrt{2}}{3} \pi \left( \frac{\ell_4}{\ell_4^{\rm (0)}}\right)^5. 
	\end{equation}
	So the knowledge of free energy teaches us the dimensionless ratio $\ell_4/\ell_4^{(0)}$ and thus the map between our conventions. Some of our comparisons are shown in Table \ref{tbl:holoF}, which includes the new solution dubbed $G_2'$ in \cite{Duboeuf:2024tbd}. Based on this, we can check numerically that $G_2'$ presented in \cite{Duboeuf:2024tbd} is identical to the one we find in section \ref{subsec:IntegralCurvesBranch2}. 
	
		\renewcommand{\arraystretch}{2}
	\begin{table}[!h]
		\centering 
		\caption{\rm We collect solutions discussed in \cite{Duboeuf:2024tbd}. The second column is the holographic free energy known in the literature or calculated here. The third column is the AdS length scale $\ell_4$ given in Table 2 of \cite{Duboeuf:2024tbd}. The third column is the corresponding quantity $\ell_4^{(0)}$ evaluated from \eqref{eq:holoF}. }
		\label{tbl:holoF}
		\begin{tabular}{ccccc}
			\hline 
			& $\frac{F_{\rm holo}}{N^{3/2}}$ & $ \ell_4$ &$ \ell_4^{(0)}$ & $\ell_4 / \ell_4^{(0)}$ \\ \hline 
			SO(8) & $\frac{\sqrt{2}\pi}{3} \approx 1.4810 $  & $\frac{1}{2}$ & $\frac{1}{2}$ & 1 \\
			SO(7)$_+$ & $\frac{2^{1/2}}{5^{3/4}} \pi \approx 1.3287 $ & 0.489270 & $\frac{1}{2}$ & $\frac{3^{1/5}}{5^{3/20}} \approx 0.97854$ \\
			SO(7)$_-$ & $\frac{2^{9/2}}{5^{5/2}}\pi \approx 1.2716$ & 0.497590 & 0.512989 & $\frac{2^{4/5}3^{1/5}}{5^{1/2}} \approx 0.96998$ \\
			$G_2$ & $ \frac{5^{5/2}}{2^2 3^{13/4}} \pi \approx 1.2356 $ & 0.489049 & 0.507092 & $ \frac{5^{1/2}}{2^{1/2}3^{9/20}} \approx 0.9644$ \\
			$G_2'$ & 1.45669 & $0.504244$ & 0.505913 & 0.996701 \\ \hline 
		\end{tabular}
	\end{table}
	\renewcommand{\arraystretch}{1}

		\begin{figure} [!h]
		\centering 
		\caption{\rm We present the $G_2''$ solution in our notation throught the map \eqref{eq:Mapping} written in coordinate $\varphi$. } 
		\includegraphics[width=4.5cm]{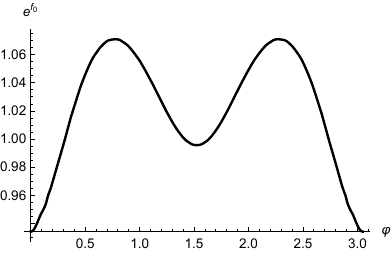} \hspace{0.3cm}
		\includegraphics[width=4.5cm]{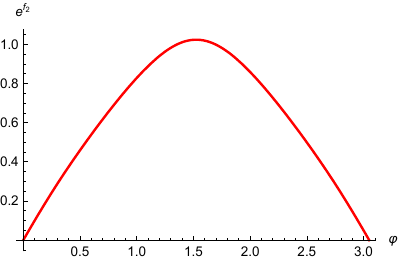} \hspace{0.3cm}
		\includegraphics[width=4.5cm]{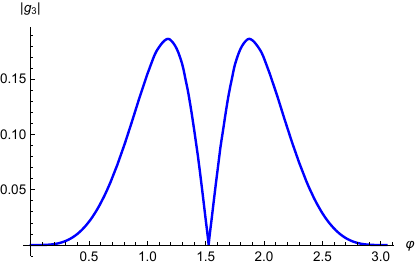}
		\label{fig:G2pp}
	\end{figure}
	Now we move on to the $G_2''$ solution. Since we don't know the correct trombone symmetry convention, we na\"ively take $\ell_4/\ell_4^{(0)} = 1$ and tranform their expression in the $\varphi$ coordinate where $f_1 = 0$,\footnote{ Notice that $e^{f_0}$ is close to a constant, we expect the solution shown in the plot is qualitatively the same as in our convention where $e^{f_1} \propto e^{f_0}$. } i.e., 
	\begin{equation}
		d\varphi = e^{f_1} d\theta.
	\end{equation}
	The result is shown in Fig. \ref{fig:G2pp}. The solution also looks smooth with a compact internal space, but the behavior of $e^{f_0}$ at small $\theta$ differs from what we expect from section \ref{subsec:theta=0}, where $e^{f_0}$ decreases instead of increasing. This suggests that while $G_2'$ is supersymmetric and preserves 4 supercharges, $G_2''$ is non-supersymmetric.

\bibliography{AdS_G2}
\bibliographystyle{JHEP}

\end{document}